# Landau quantization of Dirac fermions in graphene and its multilayers

Long-jing Yin (殷隆晶), Ke-ke Bai (白珂珂), Wen-xiao Wang (王文晓), Si-Yu Li (李思宇), Yu Zhang (张钰), Lin He (何林)[‡]

The Center for Advanced Quantum Studies, Department of Physics, Beijing Normal University, Beijing 100875, China

Corresponding author. E-mail: [‡]helin@bnu.edu.cn

When electrons are confined in a two-dimensional (2D) system, typical quantum–mechanical phenomena such as Landau quantization can be detected. Graphene systems, including the single atomic layer and few-layer stacked crystals, are ideal 2D materials for studying a variety of quantum–mechanical problems. In this article, we review the experimental progress in the unusual Landau quantized behaviors of Dirac fermions in monolayer and multilayer graphene by using scanning tunneling microscopy (STM) and scanning tunneling spectroscopy (STS). Through STS measurement of the strong magnetic fields, distinct Landau-level spectra and rich level-splitting phenomena are observed in different graphene layers. These unique properties provide an effective method for identifying the number of layers, as well as the stacking orders, and investigating the fundamentally physical phenomena of graphene. Moreover, in the presence of a strain and charged defects, the Landau quantization of graphene can be significantly modified, leading to unusual spectroscopic and electronic properties.

**PACS numbers**

## Contents

# 1  Introduction

Graphene—a one-atom-thick film—consists of a honeycomb-like hexagonal carbon lattice that exhibits a truly two-dimensional (2D) nature [1–3]. Two equivalent carbon atoms—*A* and *B*—exist in each unit cell of the hexagonal lattice in graphene [Fig. 1(a)]. These atoms are strongly connected by in-plane covalent bonds, i.e., the $\sigma$ bond hybridized between one *s* orbital and two *p* orbitals [4, 5]. The remaining unhybridized *p* orbital of each atom, which is perpendicular to the planar surface, can support the fourth valence electron of carbon, leading to the formation of a $\pi$-electronic band. The electrons filling the $\pi$ bands can move freely in the graphene plane, as if they are relativistic particles [6, 7]. These $\pi$-electronic states are responsible for the electronic properties of graphene at low energies, whereas the $\sigma$-electronic states form energy bands far from the Fermi energy. In the low-energy range, the freely moving $\pi$ electrons in graphene are described by the Dirac equation [8, 9], rather than the usual Schrödinger equation, because of the two-sublattice crystal structure. The electronic hopping between the neighboring sublattices leads to the formation of a conical energy spectrum, with the valence band and the conduction band touching at a point-like Fermi surface called the Dirac point, which yields two inequivalent points *K* and *K*′ at the corners of the hexagonal Brillouin zone (BZ) [Figs. 1(b) and 1(c)]. As a result, the quasiparticles in graphene exhibit the linear dispersion relationship $E = \hbar k v_F$ in the vicinity of the Dirac point [Fig. 1(d)], where the carriers behave as massless fermions that mimic the physics of quantum electrodynamics with a constant Fermi velocity $v_F \approx 10^6$ m/s [10, 11].

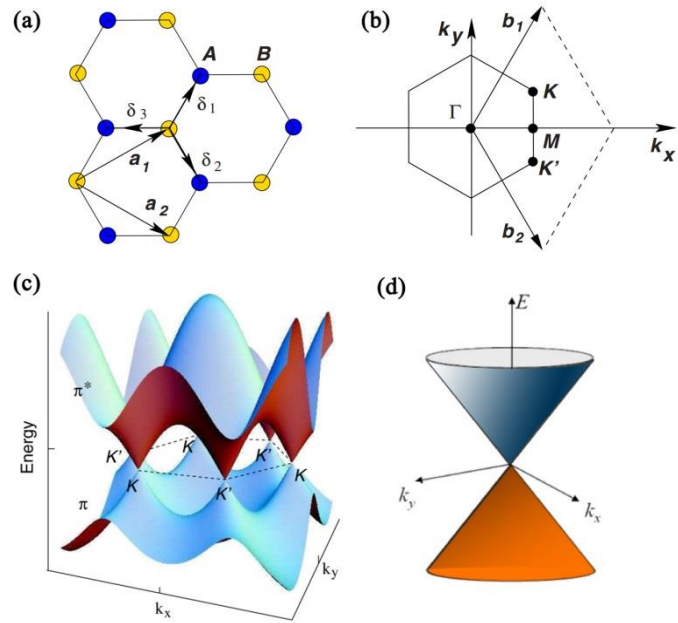

**Fig. 1** Lattice and low-energy band structures of graphene. **(a)** Honeycomb lattice consisting of *A* and *B* sublattices. **(b)** BZ. **(c)** Energy dispersion. The Dirac cones are located at the points *K* and *K'*. **(d)** The linear band structure near the Dirac point. Reproduced from Refs. [4] and [5].

In addition to the special linear spectrum, graphene has other essential and unique electronic properties that are absent from conventional 2D electron gas systems (2DEGs). For instance, because of the bipartite honeycomb lattice, the wave function for a unit cell of graphene can be expressed with two components as follows: $\Psi = a\Psi A + b\Psi B$, where $\Psi A$ and $\Psi B$ are the wave functions at the *A* and *B* sublattices, respectively. The two-component representation for graphene is very similar to that of the spinor wave functions in quantum electrodynamics. The "spin" index, which is defined by the vector (*a*, *b*), indicates sublattices rather than the real spin of electrons and is usually referred to as pseudospin. Normally, electrons and holes are not connected and are described by separate Schrödinger equations in condensed-matter physics. However, owing to the two-component wave functions of the quasiparticles in graphene, the negatively and positively charged states are interconnected, exhibiting an extra electron–hole symmetry. Additionally, the electron with energy -*E* and the hole with energy *E* originating from the band branch of the same sublattice propagate in opposite directions. Consequently, the pseudospin is parallel to the momentum for electrons and antiparallel for holes in the same branch. This introduces the chirality of the Dirac fermions in graphene [6, 12], which is positive and negative for electrons and holes, respectively. In quantum electrodynamics, the chirality is defined as the projection of spin in the direction of momentum [13]. For graphene, this definition can be used, but the true spin is replaced by the sublattice pseudospin. The 2D chiral massless-Dirac-fermions give rise to the most interesting aspects of graphene, offering a perfect platform for exploring its amazing physical phenomena, such as chiral Klein tunneling [6, 12, 14, 15], exceptional ballistic transport [16–18], and the self-similar Hofstadter butterfly spectrum [19–24].

Another interesting feature of Dirac fermions is their unusual behaviors under magnetic fields [5, 25], which lead to various novel quantum phenomena, including the anomalous integral and fractional quantum Hall effects (IQHE and FQHE, respectively) [26–31]. In the presence of a magnetic field (*B*), the low-energy band structure of a 2D electron system develops into discretely dispersionless Landau levels (LLs) [32–34], which is responsible for the IQHE [25] and is also the theoretical basis for understanding the involved FQHE [35–38]. The cyclotron energies of Dirac fermions in graphene are scaled as $\sqrt{B}$, in contrast to the linear behavior for particles in semiconductor 2DEGs. This leads to large cyclotron energies, which, together with the small scattering [39], allows the IQHE to be observed at room temperature in graphene [40]. Furthermore, there is an additional level at zero energy (i.e., at the Dirac point) including both electron and hole states [41], and accordingly, the IQHE shows an unusual half-integer characteristic.

The fully exposed electronic states provide unprecedented opportunities to directly probe 2D quantum phenomena, such as the Landau quantization, on the sample surface, not only in monilayer graphene but also in multilayer graphene [42]. In multilayer graphene, the crystal sheets are stacked on top of each other and are connected by weak van der Waals forces with different layer-stacking orders [4, 43]. The 2D nature of the quasiparticles even exists in four-layer graphene [44]. More importantly, the quasiparticles in graphene multilayers exhibit a strong layer number [45] and stacking-order dependency [46], leading to the distinct Landau quantized behaviors and rich quantum Hall physics [47–51]. This makes it possible to identify the number of layers and the stacking configurations of graphene by using the LL spectrum. One of the most powerful techniques for probing quantum phenomena such as LLs is scanning tunneling microscopy (STM) [52–54]. In scanning tunneling spectroscopy (STS) measurements, the LLs appear as a sequence of peaks in differential-conductance spectra [i.e., spectra of *dI*/*dV*, which is proportional to the local density of states (DOS)] and have been observed in many systems, including graphene [55–58], conventional 2DEGs [59, 60], and topological insulators [61–64]. This method is a local and harmless way to observe the abundant microscopic physics in graphene systems.

In this article, we review the recent experimental investigations of the Landau quantization of Dirac fermions in monolyer and multilayer graphene using STM and STS. This review is organized as follows. In Section 2, we introduce the distinct LL spectra for monolayer graphene, Bernal-stacked bilayers, and Bernal-stacked trilayers. Section 3 focuses on the experimental imaging of the two-component Dirac-LLs and the LL bending in a gapped graphene monolayer and bilayer, respectively. In Section 4, the stacking order-dependent Landau quantization in multilayer graphene is described in detail. Section 5 discusses the unconventional Landau quantized behaviors in strained and defective graphene. Finally, in Section 6, we summarize and conclude the article and present prospects for fully understanding the nature of novel quantum Hall phases and detecting new physics in graphene.

## 2  Landau quantization in graphene monolayer, Bernal bilayer, and Bernal trilayer

The truly 2D nature of quasiparticles not only exists in monolayer graphene but also extends to stacked layers of graphene sheets. Because of the exrta degree of freedom of the layer, graphene and its multilayers exhibit complex phenomena and unusual properties [65]. The most energy-stable multilayer graphene is Bernal-stacked (or AB-stacked) graphene, wherein one set of the sublattice atoms of the top layer is immediately above the atoms of the bottom layer and the other set is at the centers of the hexagonal voids in the bottom layer [66]. In graphene monolayers, the interaction of the two equivalent sublattices results in a linear band structure and gives the quasiparticles a massless Dirac fermion property near the charge neutrality with a chiral degree of $l = 1$ and a Berry phase of $\pi$. For a Bernal (AB-stacked) bilayer, the low-energy electronic states retain the chiral Dirac characteristic but have a quadratic dispersion (i.e., massive), and the quasiparticles have a chiral degree of $l = 2$ and a Berry phase of $2\pi$ [66–69]. Interestingly, Bernal trilayer (ABA) graphene exhibits the coexistence of massless and massive Dirac fermions. The various types of quasiparticles give rise to disparate Landau quantized behaviors in graphene monolayers, Bernal bilayers, and Bernal trilayers, which have been frequently observed on different substrates, especially on graphite surfaces [45, 58, 70, 71]. Graphite—a three-dimensional (3D) allotrope of carbon—consists of stacks of graphene layers with Bernal layer-stacking that are weakly coupled by van der Waals forces [43]. The surface monolayer/few-layer graphene may electronically decouple from graphite when the separation between the topmost graphene layers and the graphite is larger than the equilibrium distance of ~0.34 nm [45, 57, 72] or when the sheets have a large rotation angle with respect to the substrate after exfoliation [73]. Hence, it is possible to detect the Landau quantization of 2D Dirac fermions for graphene monolayers and multilayers on the surfaces of such 3D systems [74]. Here, we introduce the distinct LL spectra for decoupled graphene monolayers, Bernal bilayers, and Bernal trilayers on graphite and Rh substrates, which were obtained by STS under high magnetic fields.

### 2.1  Graphene monolayer

Figure 2(a) shows a series of differential-conductance spectra for a graphene monolayer on a graphite surface, under various magnetic fields ranging from 0 to 8 T [45]. The corresponding atomic-resolution STM image shows the typical honeycomb lattice structure [Fig. 2(e)]. The LLs developed as the magnetic-field increased, and the LL peaks were well-resolved up to $n = 5$ in both electron and hole sectors (here, $n$ is the LL index). Moreover, a pronounced DOS peak appeared in the vicinity of the Fermi energy under various fields, corresponding to the landmark zero-energy LL of graphene. For massless Dirac fermions in a graphene monolayer, the LL energies $E_n$ depend on the square root of both the level index $n$ and the magnetic field $B$, including a field-independent $E_0$ for the zero-energy state [55–57]:

$$E_n = \text{sgn}(n)\sqrt{2e\hbar v_F^2 |n| B} + E_0, \qquad n = 0, \pm 1, \pm 2... \qquad (1)$$

where $e$ is the electron charge, $\hbar$ is Planck's constant, $v_F$ is the Fermi velocity, and $E_0$ is the energy of the Dirac point. The unusual appearance of a zero-energy level ($n = 0$) is a direct consequence of quasiparticle chirality. Each LL in graphene is fourfold degenerate, including the zero-energy state, owing to the usual spin degeneracy and the two inequivalent corners of the BZ $K$ and $K'$, which gives rise to valley degeneracy. Theoretically, each filled single-degenerate LL contributes one conductance quantum $e^2/h$ towards the Hall conductivity observed in the quantum Hall effect (QHE) [32]. In monolayer graphene, the Hall conductivity is described by $\sigma_{xy} = (4N + 2) e^2/h$ with the absence of a zero-$\sigma_{xy}$ plateau [26, 27], in contrast to the case of conventional 2DEGs. The analysis of the data shown in Figs. 2(b) and 2(c) demonstrates that the sequence of observed LLs is described quite well by Eq. (1). The linear fit of the experimental results of the LL energies to Eq. (1), which shows electron–hole symmetry, yields a Fermi velocity of $v_F = (1.207 \pm 0.002) \times 10^6$ m/s [45]. Similar LL spectra of graphene monolayers were observed on SiC substrates [55, 56] and on Rh foil [75, 76], with the Fermi velocity of $v_F \approx 1.1 \times 10^6$ m/s in both cases.

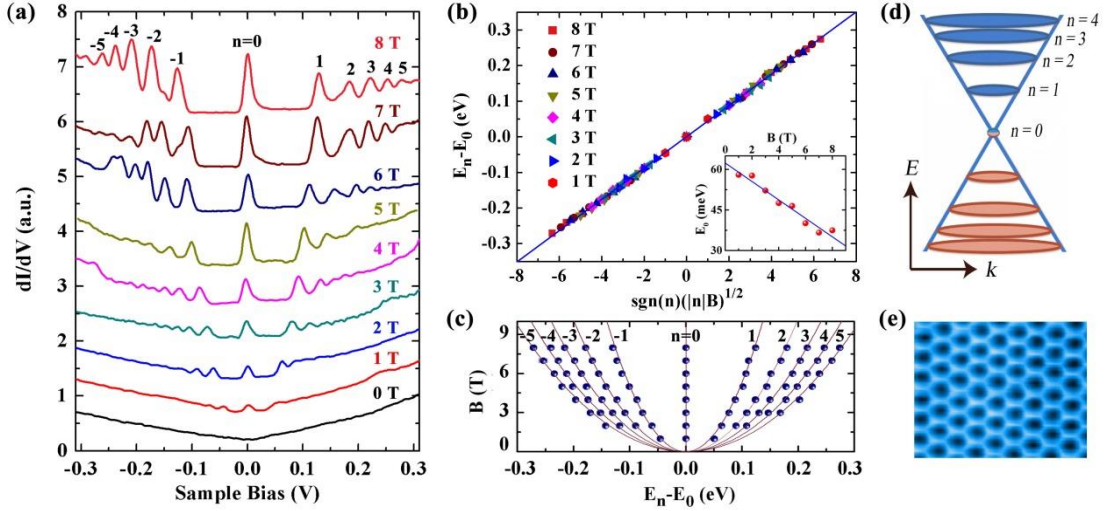

**Fig. 2** Landau quantization in monolayer graphene. **(a)** LL spectra of a graphene monolayer on a graphite surface under various fields $B$. The LL indices are marked. All the spectra are shifted to make the $n = 0$ LL remain at the same bias. **(b)** LL peak energies for 1–8 T versus $\text{sgn}(n)(|n|B)^{1/2}$. The solid line is a linear fit of the data with Eq. (1). Inset: energies of the $n = 0$ LL at different $B$. **(c)** Energies of the LL peaks for different levels $n$ show the square-root dependence on the field $B$. The solid lines are the fits with Eq. (1). **(d)** Schematic of LLs in monolayer graphene. **(e)** Atomic-resolution STM image with the honeycomb lattice structure of single-layer graphene. Reproduced from Ref. [45].

In the presence of strong magnetic fields, the electrons are more spatially localized, and the electron–electron interaction is expected to be enhanced [5]. The enhanced interaction lifts the LL degeneracies and generates gaps in graphene [56, 77], which can be directly probed in the LL spectra. In the monolayer graphene grown on Rh foil [78], we find that the energy splitting of the $LL_0$ is ~5.5 meV under a 5-T field and increases to ~8.8 meV under a 7-T field. Similar energy splits are observed in $LL_{-1}$ and $LL_1$ under a 7-T field, as shown in Fig. 3. The energy splitting of LLs with higher orbital indices is not observed in our experiment because the line width of the LLs increases with the energy. This behavior is related to the quasiparticle lifetimes, which decrease with the increasing energy difference from the Fermi level [56, 57]. Fitting the splitting energies of the $LL_0$ to a Zeeman-like dependence, $E = g\mu_B B$, yields a $g$-factor of $g \approx 21$. We attribute this energy splitting to the lifting of the valley degeneracies because the effective $g$-factor of the valley splitting in graphene is measured to be ~18.4 [56]. The observation of interaction-driven gaps, which increase with the magnetic field, clearly indicates that the electron–electron interaction in graphene is enhanced as the field increases.

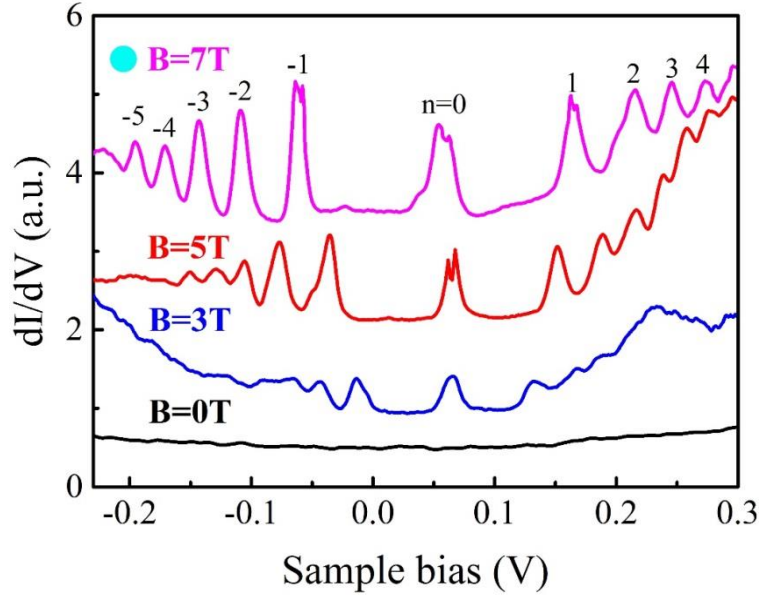

**Fig. 3** Tunneling spectra of monolayer graphene on Rh foil under different magnetic fields. The LL peak indices are labeled. The slight splitting of $LL_0$ and $LL_{\pm 1}$ is observed under a 7-T field. Reproduced from Ref. [78].

### 2.2 Bernal-stacked bilayer

Usually, in Bernal (AB-stacked) bilayer graphene, the *A/B*-atom asymmetry generated by the two adjacent AB-stacked layers leads to a triangular lattice, which is observed via microscopy. Therefore, Bernal bilayers (as well as multilayers) exhibit triangular contrast rather than the honeycomb lattice, as shown in the STM image [45, 72]. The bright spots of the triangular lattice, as shown in the inset of Fig. 5(a), are the sites on the top layer where one sublattice lies above the center of the hexagons in the second layer. In neutral Bernal bilayers, the LL spectrum of the massive Dirac fermions takes the form

$$E_n = \pm \hbar \omega_c \sqrt{n(n-1)}, \qquad n = 0,1,2... \qquad (2)$$

where $\omega_c = eB/m^*$ is the cyclotron frequency, and $m^*$ is the effective mass of the quasiparticles. This LL sequence is linear in $B$, similar to the standard 2DEGs, but has an extra zero-energy LL, which is independent of $B$. For orbital index $n > 1$, the LLs are fourfold degenerate, similar to those in monolayer graphene, while the $n = 0$ and $n = 1$ LLs are further degenerate, resulting in an eightfold degenerate zero-energy state (Fig. 4). Therefore, the Hall conductivity in bilayer graphene exhibits plateaus at integer values of $4e^2/h$ and has a double $8e^2/h$ step between the hole and electron states across zero density [47, 48, 79]. However, in the supported bilayers, the substrate easily induces an interlayer bias, breaking the inversion symmetry of the two adjacent layers [77, 80]. Consequently, the degeneracy of the zero-energy LL is partially lifted, and a bandgap is opened in the low-energy bands [81–84]. Then, there are two DOS peaks located at the edges of the energy gap ($E_g$) in the *dI/dV* spectra, even under zero field, as shown in Fig. 5. With an increasing magnetic field, the DOS peaks at the gap edges become two valley-polarized quartets, i.e., $LL_{(0,1,+)}$, and $LL_{(0,1,-)}$ [85], which are almost independent of $B$, and the spectra develop into a sequence of well-defined LL peaks showing a linear field dependence. These features are the fingerprints of the quantized behavior of massive Dirac fermions in gapped graphene bilayers.

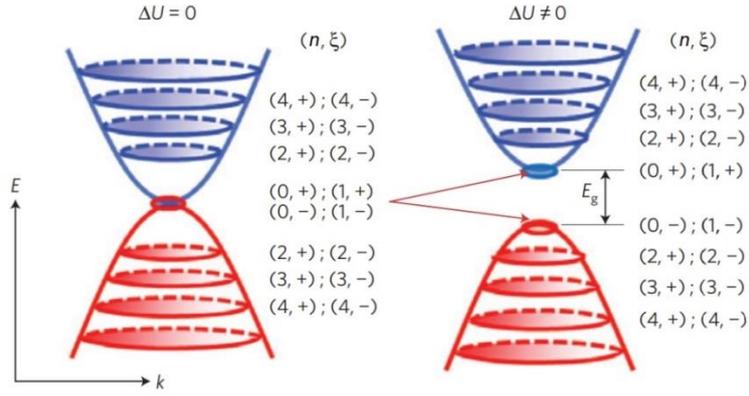

**Fig. 4** Schematic of Landau quantization in bilayer graphene with and without an energy gap. LLs are indexed by the orbital and valley indices: $n$ and $\xi$. The eightfold-degenerate zero-energy LL splits into two valley-polarized quartets under an interlayer bias. Reproduced from Ref. [77].

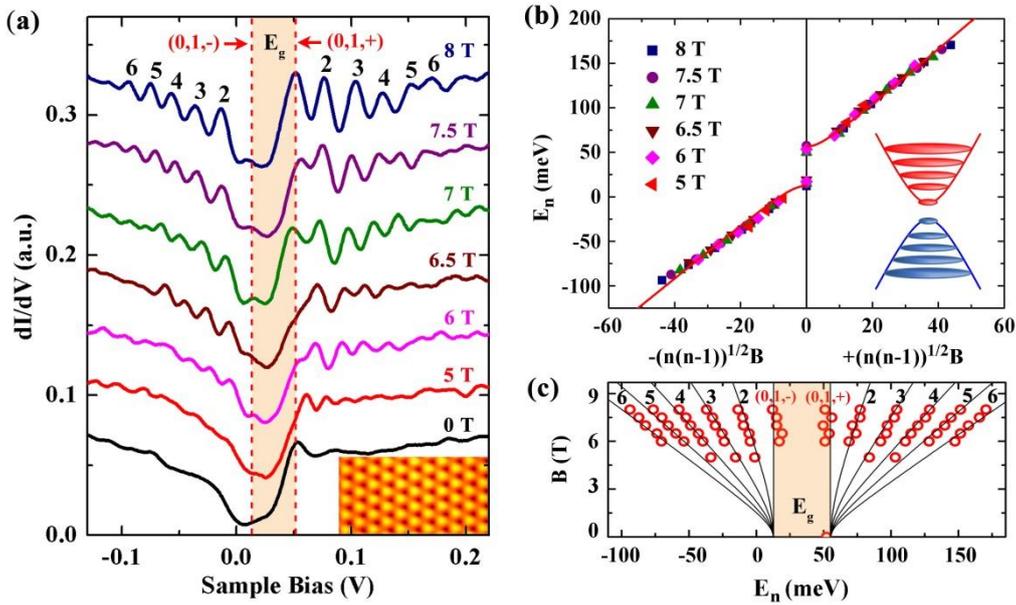

**Fig. 5** Landau quantization in bilayer graphene. **(a)** LL spectra of a graphene bilayer on a graphite surface under various fields $B$. The LL indices are labeled. Inset: atomic STM image of a bilayer showing a triangular lattice. **(b)** LL peak energies plotted against $\pm(n(n-1))^{1/2} B$. The solid lines are the fits of the data with Eq. (3). Inset: schematic of the LLs in bilayer graphene with a finite gap. **(c)** Fan diagram of the LLs energies as a function of the magnetic field. The solid lines represent fits of the data with Eq. (3). Reproduced from Ref. [45].

For a gapped graphene bilayer, the LL sequences can be described as follows [86, 87]:

$$E_n = E_C \pm \sqrt{(\hbar\omega_c)^2[n(n-1)] + (U/2)^2} - \xi z U/4, \qquad n = 2,3,4... \qquad (3)$$
$$E_0 = E_C + \xi U/2, \qquad E_1 = E_C + \xi(U/2)(1-z),$$

where $E_C$ is the energy of the charge-neutrality point (CNP), $U$ is the interlayer bias, and $\xi = \pm$ are valley indices. Normally, $z = 2\hbar\omega_c/t_\perp \ll 1$ for $B \leq 8$ T and $|U| \approx E_g$ when $U < t_\perp$. According to the fit between the experimental data and Eq. (3) shown in Figs. 5(b) and 5(c), we can determine $E_g$ and $m^*$. For the bilayer shown in Fig. 5, $E_g \approx 40$ meV and $m^* = (0.039 \pm 0.002)m_e$ ($m_e$ is the free-electron mass). We observed various bilayer graphene samples on the graphite substrate, with $E_g$ ranging from 10 to 100 meV and $m^*$ of 0.03–0.05$m_e$. Both the effective mass of the massive Dirac fermions and the bandgap agree well with the range of values reported previously for Bernal graphene bilayers on different substrates [67, 68, 77, 88]. In *AB*-stacked bilayers, the wave functions for one valley (+) of the lowest LL are mainly localized on the *B* sites of the top layer, and the wave functions for the other valley (-) of the zero-energy LL are on the *A* sites of the bottom layer. Therefore, the signal of LL$_{(0,1,+)}$ (mainly localized on the first layer) is far stronger than that of LL$_{(0,1,-)}$ (localized on the second layer) in the spectra, as the STS predominantly probed the DOS of the electrons on the top layer [72, 77], as shown in Fig. 5(a). This feature can be considered as another signature of gapped bilayer graphene and can be used to directly identify the bilayer region on the nanoscale.

## 2.3 Bernal-stacked trilayer

According to tight-binding calculations that include only nearest-neighbor intralayer and nearest-layer coupling, the low-energy spectrum for Bernal trilayer can be treated as the combination of those for a graphene monolayer and a Bernal bilayer [89, 90]. Consequently, the Landau quantization of an ABA trilayer is a superposition of two sequences of massless and massive Dirac fermions [91, 92]. The *dI/dV* spectra measured in trilayer graphene under various magnetic fields exhibit a sequence of LL peaks of both massless and massive Dirac fermions, as shown in Fig. 6. In Fig. 6(a), some of the LL peaks depend on the square root of the magnetic field (massless-type), as shown in Fig. 6(b). The other LLs exhibit a linear-field dependence (massive-type), as shown in Fig. 6(c). In the trilayer sample of Fig. 6, the field dependence of the LLs corresponding to the massive Dirac fermions in both electron and hole sectors is extrapolated to the same zero-field value, suggesting that the eightfold degeneracy of the zero-energy LL in the bilayer-like sub-bands is not lifted. For convenience, in this case, the LL spectrum of the massive Dirac fermions can be expressed in the form [70]

$$E_n = \text{sgn}(n)\hbar\omega_c\sqrt{|n|(|n|+1)} + E_0, \qquad n = 0, \pm 1, \pm 2.... \qquad (4)$$

A good fit of the bilayer-like LL energies in Fig. 6(a) with Eq. (4) is shown in Fig. 6(c). Furthermore, the ABA trilayers, having parabolic sub-bands containing a finite bandgap of ~10 meV and two valley-polarized quartets under *B*, are observed in our STM measurements. However, the linear sub-bands are always gapless. As a result, Bernal trilayer graphene cannot open a bandgap under low-energy excitation [93–95]. Theoretically, the CNPs of the monolayer graphene-like and Bernal bilayer-like sub-bands in the ABA trilayers are at the same energy, according to the simplest approximation [Figs. 6(d) and 6(e)]. In the experiments, there is a slight energy difference of approximately 10–30 meV between the zero-energy states of massless and massive Dirac fermions [45]. Such a difference is very reasonable when the other non-nearest neighbor hopping parameters affect the band structure of the Bernal trilayer.

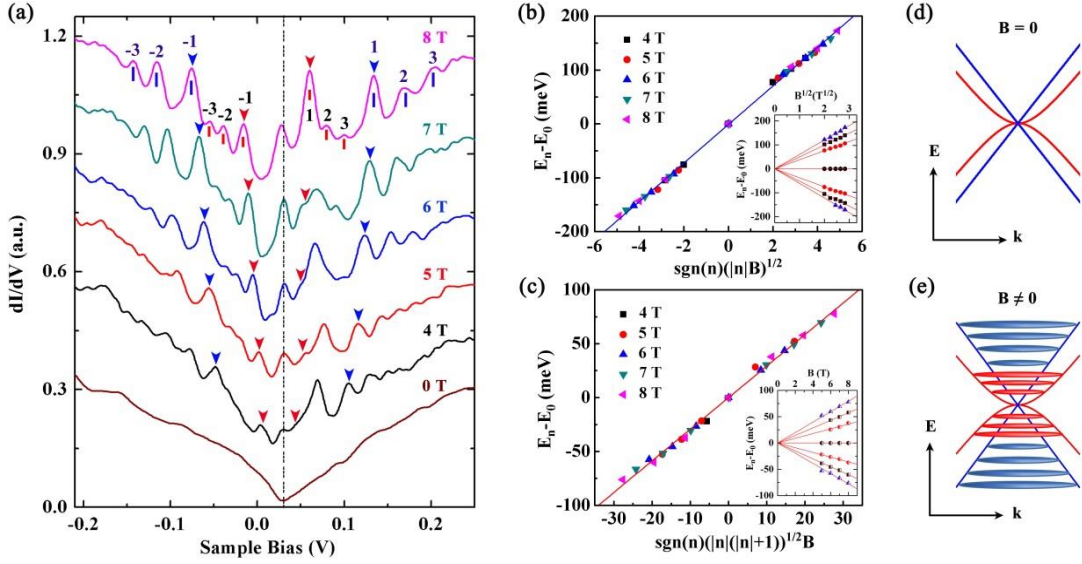

**Fig. 6** Landau quantization in Bernal trilayer graphene. **(a)** LL spectra of a graphene trilayer on a graphite surface under various fields $B$. The LL indices of massless and massive Dirac fermions are indicated by blue and black numbers, respectively. **(b, c)** LL peak energies of massless and massive Dirac fermions versus $\mathrm{sgn}(n)(|n|B)^{1/2}$ and $\mathrm{sgn}(n)(|n|(|n|+1))^{1/2}B$, respectively. Schematic low-energy band structure of the graphene trilayer around the point $K$ under $B = 0$ **(d)** and $B \neq 0$ **(e)**. Reproduced from Ref. [45].

The Fermi velocity of massless Dirac fermions and the effective mass of massive Dirac fermions in ABA trilayers can be obtained by fitting the LL peaks to corresponding theoretical formulas, as shown in Figs. 6(b) and 6(c). Interestingly, the value of $v_F$ in different Bernal trilayers can differ by more than 30%, similar to that observed in different graphene monolayers, which ranges from $0.79 \times 10^6$ to $1.21 \times 10^6$ m/s [55–57]. Additionally, a strong correlation between $v_F$ and $m^*$ is observed: $v_F$ generally increases as $m^*$ decreases, as shown in Fig. 7(a). Thus, there is a common origin that simultaneously affects $v_F$ and $m^*$ in the ABA trilayers. This behavior can be explored by calculating the band structure of the Bernal trilayer with different hopping parameters, as $v_F$ and $m^*$ depend sensitively on the hopping parameters. For example, the increase in the nearest-neighbor intralayer coupling in the ABA graphene can lead to opposite variations in $v_F$ and $m^*$; i.e., $m^*$ decreases while $v_F$ increases, as shown in Fig. 7(b). This effect can quantitatively explain the experimental results.

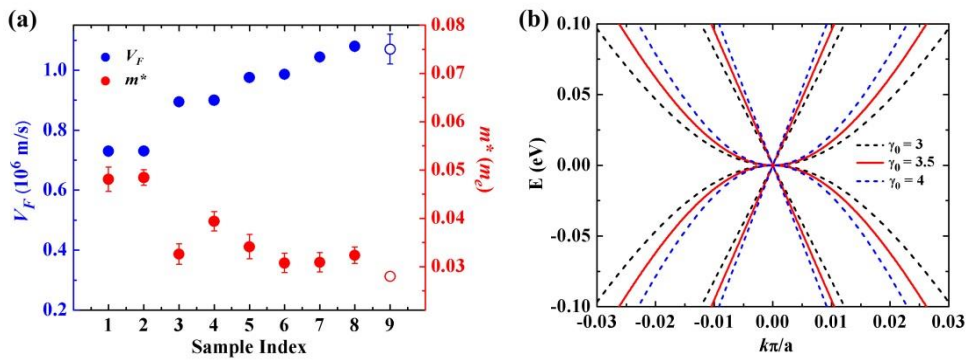

**Fig. 7 (a)** The $v_F$ of massless Dirac fermions and the $m^*$ of massive Dirac fermions obtained in different Bernal trilayers. **(b)** Low-energy band structure of a Bernal graphene trilayer with different nearest-neighbor intralayer hopping strengths. Reproduced from Ref. [45].

## 3 Inherent 2D Dirac quantum properties

The 2D chiral Dirac fermion system of graphene provides a perfect platform for theoretically [96, 97] and experimentally [98–101] identifying and examining extraordinary and fundamental physical problems, which are unachievable in 3D structures. The fully surface-exposed electronic states allow the internal quantum properties of the quasiparticles in graphene to be directly imaged by microscopic methods, such as scanning-probe techniques [53, 54]. We show that by using STM and STS with high space- and energy-resolution simultaneously, the two-component characteristic of Dirac fermions and the LL bending are spatially visualized in gapped graphene monolayers and bilayers, respectively.

### 3.1 Two-component Dirac-LLs

Because of the bipartite honeycomb lattice in graphene [4–6], which has two distinct sublattices (*A* and *B*), the wave functions describing the low-energy excitations near the Dirac points in monolayer graphene are two-component spinors. This two-component nature can be manifest in localized LLs, whose degeneracy is lifted by the Coulomb potential or interactions. Therefore, with the high energy and spatial resolution of STM and STS, it is possible to image the two components of the spinors in a uniform graphene system.

#### 3.1.1 Two-component spinors

The two-component spinors of the two Dirac cones (*K* and *K'*) in graphene have the following form:

$$|K\rangle = \begin{pmatrix} \psi_{KA} \\ \psi_{KB} \end{pmatrix} = \frac{1}{\sqrt{2}} \begin{pmatrix} 1 \\ \pm i e^{-i\theta_\tau} \end{pmatrix}, \quad |K'\rangle = \begin{pmatrix} \psi_{K'B} \\ \psi_{K'A} \end{pmatrix} = \frac{1}{\sqrt{2}} \begin{pmatrix} 1 \\ \mp i e^{-i\theta_\tau} \end{pmatrix}. \quad (5)$$

Here $\theta_\tau = \arctan\left(\frac{q_{\tau,y}}{q_{\tau,x}}\right)$ is defined as the angle of the wave vector $\boldsymbol{q}_\tau \equiv (q_{\tau,x}, q_{\tau,y})$ in the momentum space. The two-component representation, which mathematically resembles that of a spin [102], corresponds to the projection of the electron wave function on the *A* and *B* sublattices.

A site energy difference of 2Δ between the sublattices can break the inversion symmetry of graphene and lift the energy degeneracy of the *A* and *B* sublattices [96, 103], as shown in Fig. 8(a). This generates a gap of Δ*E* = 2Δ at the Dirac points, as shown in Fig. 8(b), which was observed for a graphene monolayer on top of SiC [88], graphite [30, 57, 104], and hexagonal boron nitride [19, 105]. The gap, which usually ranges from 10 meV to several tens of millielectron volts, can result in valley-contrasting Hall transport in graphene monolayers [96, 105]. In the quantum Hall regime, the broken symmetry of the graphene sublattices shifts the energies of the *n* = 0 LL in the *K* and *K'* valleys in opposite directions and thus splits the *n* = 0 LL into the 0₊ and $0_-$ LLs (here, $\lambda = +$ and $-$ denote the $K'$ and $K$ valleys, respectively) [57, 104], as schematically shown in Fig. 8(c). Generally, the wave functions $|n_\lambda\rangle$ of the LLs in graphene are given by [5, 106, 107]

$$|n_-\rangle = \begin{pmatrix} \psi_A^{n_-} \\ \psi_B^{n_-} \end{pmatrix} = \begin{pmatrix} \sin\alpha_{n_-} \phi_{|n|-1} \\ \cos\alpha_{n_-} \phi_{|n|} \end{pmatrix}, \quad |n_+\rangle = \begin{pmatrix} \psi_B^{n_+} \\ \psi_A^{n_+} \end{pmatrix} = \begin{pmatrix} \cos\alpha_{n_+} \phi_{|n|-1} \\ \sin\alpha_{n_+} \phi_{|n|} \end{pmatrix}, \quad (6)$$

where $\tan\alpha_\pm = \left(\hbar\omega_B\sqrt{|n|}\mathrm{sgn}(n)\right)/\left(\sqrt{\hbar^2\omega_B^2|n| + \Delta^2} - \Delta\mathrm{sgn}(n)\right)$ $(n \neq 0)$, and $\omega_B = \sqrt{2ev_F^2B/\hbar}$, with $v_F$ being the Fermi velocity and $B$ being the magnetic field (here, $\phi_n$ is the usual LL wave function). For $n = 0$, $\alpha_{0_-} = 0$ and $\alpha_{0_+} = \pi/2$; hence, only the second components of the spinors are nonzero, and we have $|0_-\rangle = \begin{pmatrix} \psi_A^{0_-} = 0 \\ \psi_B^{0_-} = \phi_0 \end{pmatrix}$ and $|0_+\rangle = \begin{pmatrix} \psi_B^{0_+} = 0 \\ \psi_A^{0_+} = \phi_0 \end{pmatrix}$. This indicates that we can detect the $0_-$ LL, i.e., the spinor $|K_0\rangle$, only on the $B$ sites and detect the $0_+$ LL, i.e., the spinor $|K'_0\rangle$, only on the $A$ sites.

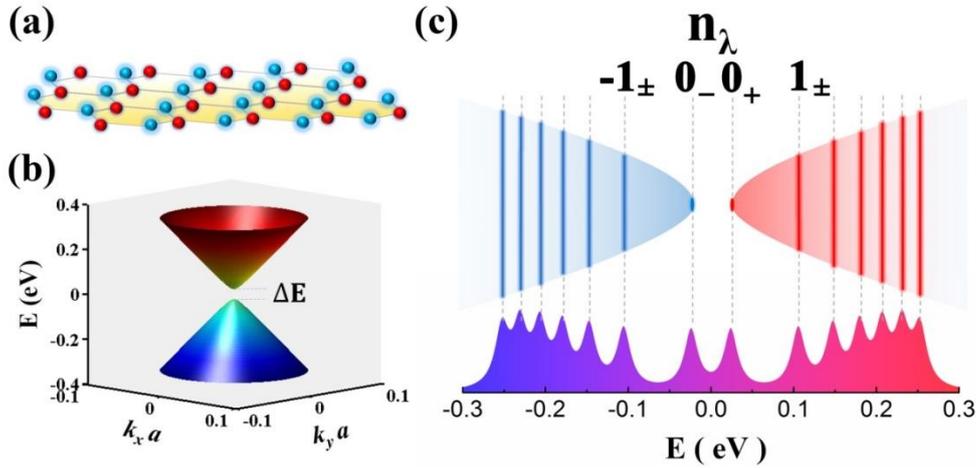

**Fig. 8** Electronic band structure and Landau quantization for a gapped graphene monolayer. **(a)** Schematic diagram of a graphene monolayer with a staggered sublattice potential breaking the inversion symmetry. The $A$- and $B$-sites are denoted by blue and red balls, respectively. **(b)** Energy spectrum of a graphene monolayer with broken inversion symmetry. **(c)** Schematic LLs and DOS of a gapped graphene monolayer in the quantum Hall regime. The peaks in the DOS correspond to the LLs $n_\lambda$. Reproduced from Ref. [108].

### 3.1.2 Localized Dirac-LLs

Figures 9(a) and 9(d) show representative STM images of the decoupled graphene layer on SiC and graphite substrates [108]. The spectra of the graphene sheets, which were recorded under a magnetic field of 8 T [Figs. 9(c) and 9(f)], on both the SiC and graphite substrates exhibit the Landau quantization of Dirac fermions, as expected for a gapped graphene monolayer [45, 55, 57]. The Fermi velocities for the graphene sheets on the SiC and graphite substrates are estimated to be $v_F = (0.79 \pm 0.03) \times 10^6$ and $(0.84 \pm 0.03) \times 10^6$ m/s, respectively. A notable feature of the tunneling spectra is the splitting of the $n = 0$ peak and its sensitive dependency on the recorded positions depicted in Figs. 9(c) and 9(f). The splitting of the $n = 0$ peak, ~20 mV, is attributed to a gap caused by the substrate potential breaking the inversion symmetry [55, 57]. The spectra recorded at different positions, as shown in Figs. 9(c) and 9(f), indicate that the $0_-$ LL is significant only on the $B$ sites and that the $0_+$ LL is pronounced only on the $A$ sites. At the center of the hexagons of the graphene sheets, for example, at the green dots in Figs. 9(d) and 9(e), the observed intensities of the $0_+$ and the $0_-$ LLs are almost identical, as shown in Figs. 9(c) and 9(f). This feature is related to characteristics of the internal structure of the two-component spinors of the $0_-$ and $0_+$ LLs. The following results demonstrate that the splitting of the $n = 0$ LL is a direct consequence of this two-component nature.

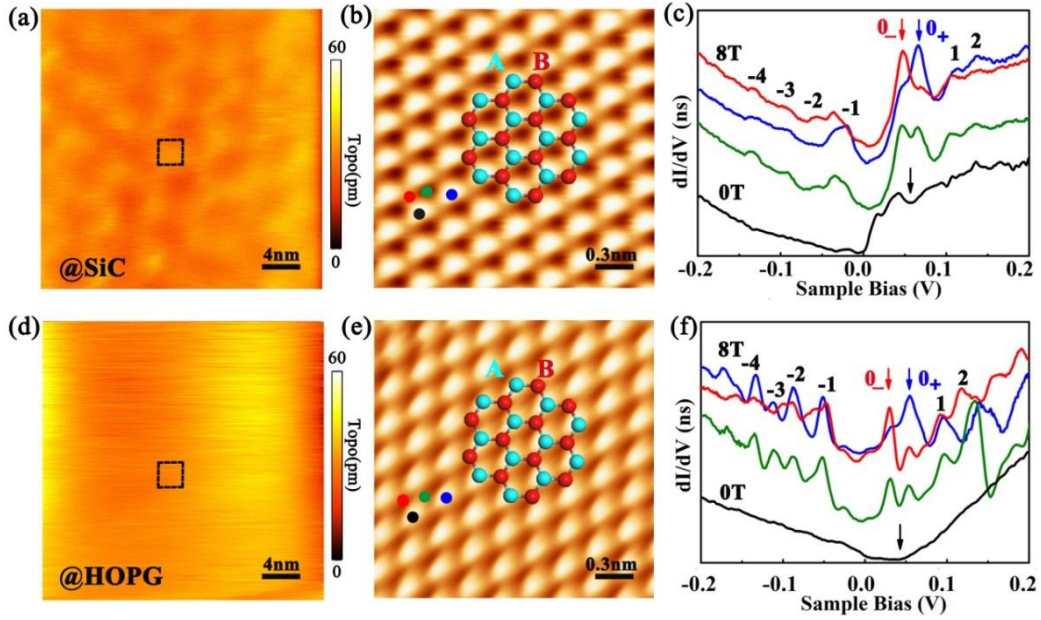

**Fig. 9** STM images and STS spectra of gapped graphene sheets. **(a, d)** STM images of graphene on a SiC (000-1) terrace and on HOPG, respectively. **(b, e)** Magnified atomic-resolution topographies for the black frames in (a) and (d), respectively. Here, the bright and dark spots represent the $A$- and $B$-site atoms, respectively. **(c, f)** $dI/dV$ spectra obtained at different positions, as indicated by the different colors, in (b) and (e), respectively. The black arrows in both panels denote the position of the CNP of the topmost graphene sheets under zero magnetic field. The LL indices are marked. Reproduced from Ref. [108].

Figures 10(a) and 10(b) show differential-conductance maps for graphene on a SiC substrate under an 8-T field at the bias voltages of the $0_+$ and $0_-$ LLs, respectively. The maps reflect the spatial distribution of the local DOS at the bias voltages. Both the maps exhibit triangular contrasting, as indicated the pronounced asymmetry of the $0_+$ and $0_-$ LLs on the sublattices. However, there is a very important difference between the two maps. The bright spots in the conductance map of the $0_-$ LL correspond to the dark spots of the triangular lattice, i.e., the $B$ sites, in the STM image, whereas the bright spots in the map of the $0_+$ LL correspond to the bright spots of the triangular lattice, i.e., the $A$ sites, in the STM image. At a fixed energy, the local DOS at the position $r$ is determined by the wave functions, according to $\rho(r) \propto |\psi(r)|^2$. Therefore, the maps shown in Figs. 10(a) and 10(b) reflect atomic-resolution images of the two-component Dirac-LLs. Theoretically, the spinor of the $0_+$ ($0_-$) LL only has a non-zero component on the $A$ ($B$) sites, which is qualitatively consistent with the observed large asymmetry of the $0_+$ and $0_-$ LLs on the sublattices.

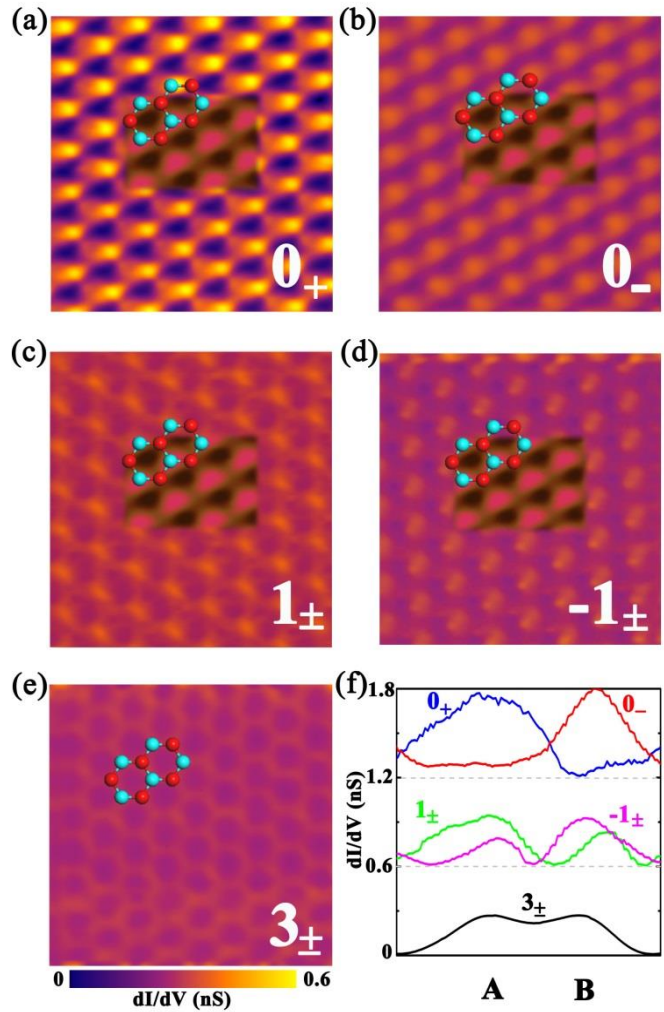

**Fig. 10** Conductance maps of the gapped graphene monolayer on SiC at different energies. **(a–e)** Conductance map recorded at the bias voltages of the $0_+$ LL ($V_b$ = 65.5 mV), $0_-$ LL ($V_b$= 45 mV), +1 LL ($V_b$ = 108 mV), -1 LL ($V_b$ = -22 mV), and +3 LL ($V_b$ = 170 mV), respectively. The honeycomb structure of graphene and the atomic-resolution STM image are overlaid onto the maps. **(f)** Vertical line-cuts of the conductance maps of the $0_+$, $0_-$, -1, +1, and +3 LLs along the *A* and *B* atoms. The curves are offset vertically for clarity, and the zero lines for these curves are denoted by dashed lines. Reproduced from Ref. [108].

Theoretically, $\sin^2(a_n/2)/\cos^2(a_n/2)$ which reflects the asymmetry between the amplitudes of the *A*-site and *B*-site components of the spinors. Figures 10(c) and 10(d) show conductance maps for graphene on a SiC substrate under an 8-T field at the bias voltages of the *n* = +1 and *n* = -1 LLs, respectively. Both the maps exhibit triangular contrasting, and the amplitude of the *n* = +1 (*n* = -1) LL on the *A* (*B*) sites is clearly stronger than that for the *B* (*A*) sites. However, the asymmetry between the amplitudes of the *A*-site and *B*-site components of the spinors for the *n* = +1 and *n* = -1 LLs is far weaker than that for the $0_+$ and $0_-$ LLs, as shown in Figs. 10(a)–10(d). This n asymmetry decreases as |*n*| increases, according to Eq. (6). Therefore, Fig. 10(e) shows almost honeycomb contrasting in the conductance map recorded at a bias voltage of 170 mV (*n* at 170 mV is estimated to be 3). The vertical line-cuts of the conductance maps of the $0_+$, $0_-$, -1, +1, and +3 LLs along the *A* and *B* atoms in Fig. 10(f) also depict the asymmetry between the amplitudes of the *A*-site and *B*-site components.

## 3.2 LL bending

Under a strong magnetic field $B$, the low-energy band structure of graphene is divided into dispersionless LLs and causes insulating behavior in the bulk material (see Fig. 11), while the confining potential at the edges [104, 109, 110] of the system bends the discrete LLs to form dispersive edge states that carry charge carriers in the QHE [111–113]. Hence, the LL bending is a fundamental effect and is significant for fully understanding the nature of the graphene edge states in novel quantum Hall phases such as the symmetry-protected quantum spin Hall state [114, 115] and exotic ferromagnetic phases [116, 117]. Because the electrons reside at the graphene surface, in contrast to the case of semiconductor-based 2DEGs, it should be possible to directly probe the level bending at the edges and perform systematic studies of the edge states for testing theoretical ideas of quantum Hall edge physics [118]. Numerous works have addressed this subject, mainly using optical and transport measurements [119–121]. Direct experimental observation of the LL bending can be achieved by STM and STS at different edge terminations of graphene on the graphite surface [72].

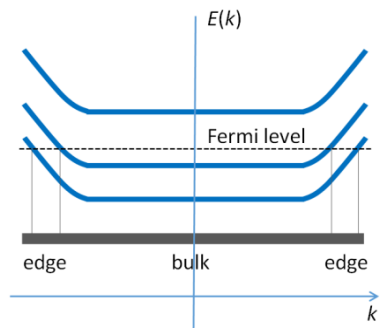

**Fig. 11** LL structures of the 2DEGs in a finite-size sample. The LLs are dispersionless in the bulk material but dispersive near the sample edges.

There are two possible (perfect) edge terminations—zigzag and armchair (see Fig. 12)—in graphene [122, 123], and the edge orientations strongly affect the electronic structures of graphene sheets [124–126]. The zigzag edge is predicted to host surface states with penetration into the bulk [127, 128] and has attracted much attention [129–132] because such surface states are believed to be closely related to the bandgap opening [128, 133, 134], magnetic order [135], and exceptional ballistic transport [16]. In the quantum Hall regime, both the zigzag and armchair edges can bend LLs to generate dispersive edge states, even in bilayer graphene [136–138]. Graphene layers such as bilayers with different edge terminations (Fig. 12) can be identified by preforming STM and STS under magnetic fields, as discussed in Section 2. Figure 12(b) shows typical and well-defined LL spectra of gapped bilayer graphene on a graphite surface under various fields $B$ [72]. The high-quality sample with atomically sharped edges and the ultralow random potential fluctuations due to substrate imperfections on the graphite allow us to directly probe the LL bending near the edges.

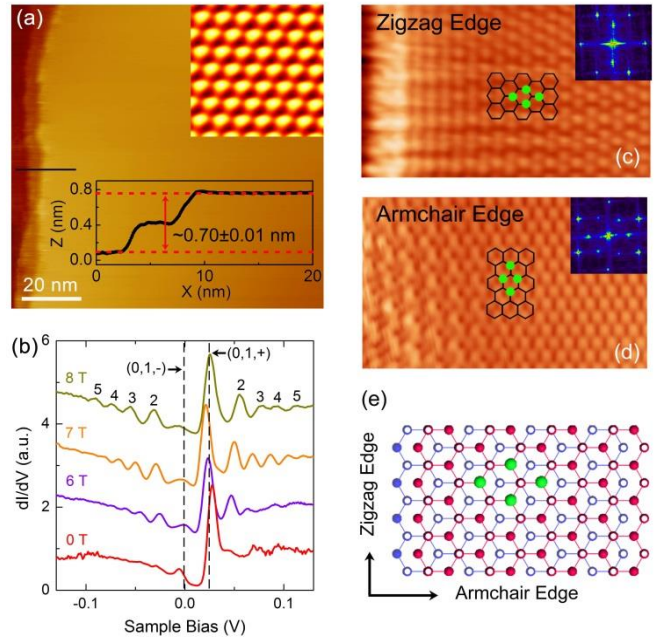

**Fig. 12** Bilayer graphene with different edge terminations. **(a)** STM topographic image of a bilayer graphene near the sample edge on a graphite surface. Inset (upper): atomic-resolution image showing the triangular contrast. Inset (lower): height profile along the black line across the edge. **(b)** STS spectra of the graphene bilayers recorded away from the edges under various fields *B*. The LL peak indices are labeled. STM atomic images of **(c)** a zigzag bilayer edge and **(d)** an armchair bilayer edge. The insets show 2D Fourier transforms of the images. **(e)** Schematic of the *AB*-stacked bilayer graphene with zigzag and armchair edges. The green dots represent a set of sublattices imaged via STM topography. Reproduced from Ref. [72].

Figure 13 summarizes the STM results measured near two bilayer graphene edges in the quantum Hall regime. The expected LL bending at both the zigzag and armchair edges are clearly observed. Away from the edges, the LL spectra follow the sequence of massive Dirac fermions in gapped bilayers [Figs. 13(a)–13(d)]. Approaching the edges, the DOS peaks for the LLs become weak, and the LLs are shifted away from the CNP, as shown in Figs. 13(e) and 13(f). The local DOS measured at position *r* with a fixed energy is determined by the wave functions according to $\rho(r) \propto |\psi(r)|^2$, while the wave functions of the LLs have a spatial extent of $\sim 2\sqrt{N}l_B$ ($l_B = \sqrt{\hbar/eB}$ is the distance from the edge) [61, 104]. Therefore, there is an important contribution from the bulk states, even for the LL spectra measured near the edges. Moreover, the wave functions of LLs with higher indices have greater spatial extents, as shown in the inset of Fig. 13(f). Consequently, the amplitude of the high-index LL peaks decreases more slowly than that of low-index LL peaks [Fig. 13(e)], and the bending of the low-index LLs seems stronger than that of the high-index LLs [Fig. 13(f)] (because of the greater contribution from the bulk states to the higher LLs). We also observed the LL bending under different magnetic fields, as shown in Fig. 14.

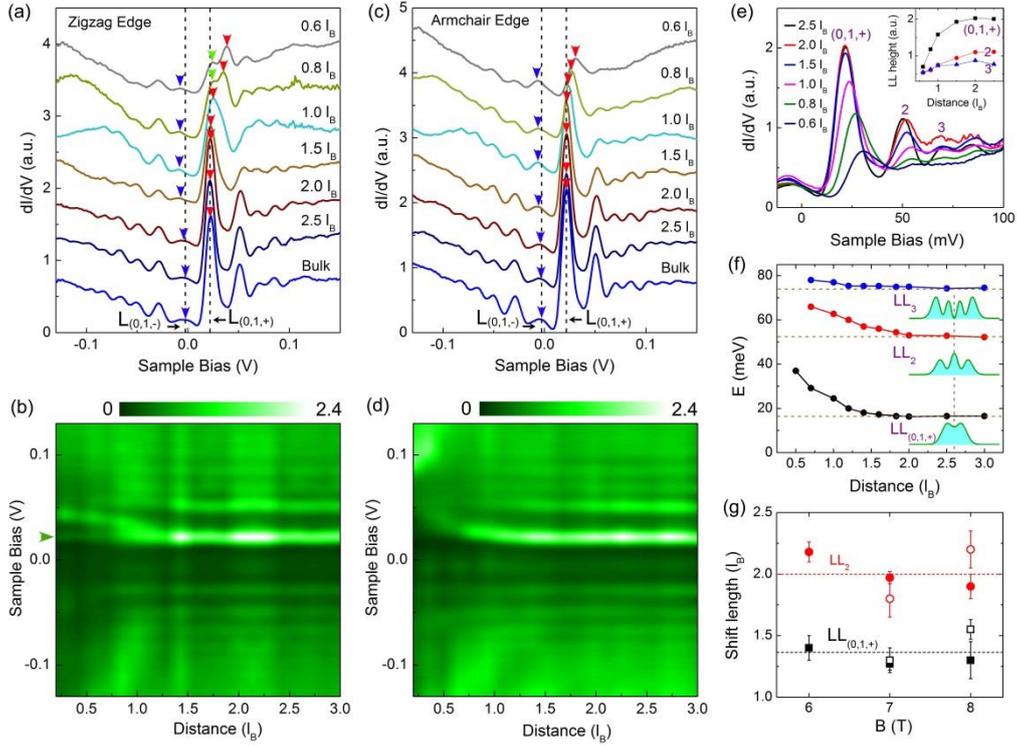

**Fig. 13** LL bending at the bilayer graphene edges. (**a, c**) Spatial variation of the LL spectra near the zigzag (a) and armchair edges (c) of bilayer graphene under a 7-T field. The blue and red arrows indicate the spatial evolution of the $LL(0,1,-)$ and $LL(0,1,+)$ peaks. (**b, d**) LL spectra maps measured near the zigzag and armchair edges, respectively, under a field of 7 T. The quasi-localized surface states of the zigzag edge are indicated by green arrows in (a) and (b). (**e**) Evolution of the LL peaks under a 7-T field near the armchair edge on the conduction-band side. Inset: LL peak heights extracted from (e) as a function of the distance from the edge. (**f**) LL bending as a function of the distance around the armchair edge under an 8-T field. Inset: calculated probability densities for the wave functions of $LL(0,1,+)$, LL2, and LL3 under an 8-T field. (**g**) Shift lengths of $LL(0,1,+)$ and LL2 under different fields $B$. The solid dots (open dots) correspond to the data for the armchair edges (zigzag edges). The dashed lines show the average values of $\sim 1.4 l_B$ and $\sim 2.0 l_B$ for $LL(0,1,+)$ and LL2, respectively. Reproduced from Ref. [72].

The shift length of the LL bending around the edges is theoretically predicted to be of the magnetic length [112, 136]. Figure 13(g) summarizes the measured shift length under different fields $B$ around both the zigzag and armchair edges. Here, we observe that the shift length depends on neither the magnetic fields nor the edge types and it is of the magnetic length. However, the shift length appears to be dependent on the orbital index: the estimated shift lengths for $LL_{(0,1,+)}$ and $LL_2$ are approximately $1.4 l_B$ and $2.0 l_B$, respectively. We obtained the average value of the bending energy for the lowest $LL_{(0,1,+)}$: ~65 meV (see Fig. 14). This energy scale is approximately identical to the depth of the confining potential well at the sample edge [137, 139].

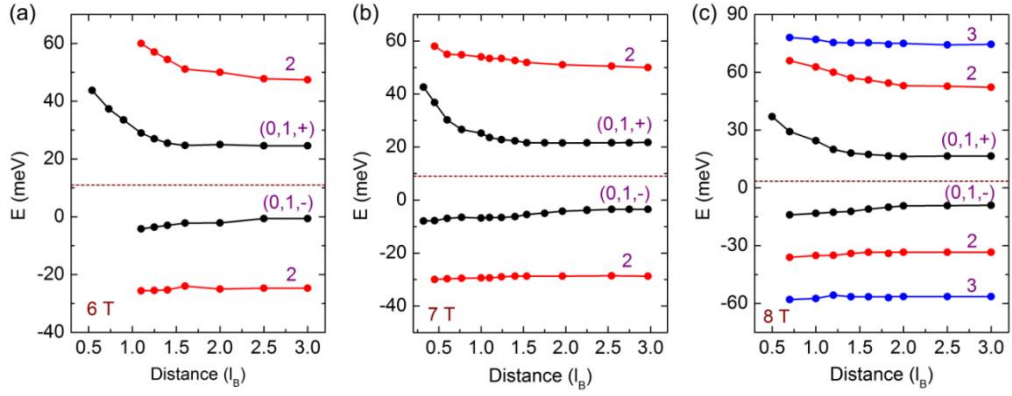

**Fig. 14** LL bending for the armchair edge measured under various fields $B$. The dashed lines are the energy positions of CNPs in the bilayers. Reproduced from Ref. [72].

## 4 Stacking-dependent LL spectrum for multilayer graphene

Although the most common multilayer graphene is AB-stacked, other natural stable layer configurations exist in the 2D crystals [66]. Structurally, to form other configurations, one graphene sheet of the multilayers can be shifted on the atomic scale along the lattice orientation (usually in the armchair direction) or twisted at a finite angle with respect to the adjacent layers. In the former case, degenerate stacking orders related to the AB-register are generated, such as BA-stacking in bilayers and rhombohedral (ABC) stacking in multilayers [140, 141]. For the latter, it introduces a stacking-misorientation structure, that is, twisted graphene, forming Moiré superlattices [142]. In both cases, the electronic properties of the graphene layers can be modified dramatically and depend sensitively on the stacking orders [46, 143–145]. In this section, we present the characteristic features of the stacking-dependent LL spectrum in graphene bilayers and trilayers.

### 4.1 Stacking domain wall system

When multilayer graphene partly transforms from one stacking order to another configuration, one-dimensional (1D) strain soliton-like domain walls grow in the transition boundary [146]. For example, the region of the transition between AB- and BA-stacked bilayer graphene domains forms an AB-BA domain wall [147–149], and similarly, graphene trilayers have ABA-ABC stacking solitons [150]. To realize such domain walls, one graphene layer should be shifted one C-C bond length along the armchair direction with respective to the adjacent layers (see Fig. 15). If the displacement vector is parallel to the domain wall, a shear-type soliton is produced [Fig. 15(a)], whereas if it is perpendicular to the domain wall, a tensile-type one is produced [Fig. 15(b)]. Owing to the varied interlayer stacking around the domain walls, the AB-BA and ABA-ABC stacking solitons can be clearly visualized by STM [73, 150], as shown in Figs. 15(c) and 15(d), by STEM [147] or by infrared nanoscopy [151]. The fascinating electronic and optical properties in these 1D solitons have been demonstrated theoretically [152–154] and experimentally [73, 151, 155, 156], including the topologically protected valley Hall edge states in AB-BA bilayer graphene domain walls [73, 155].

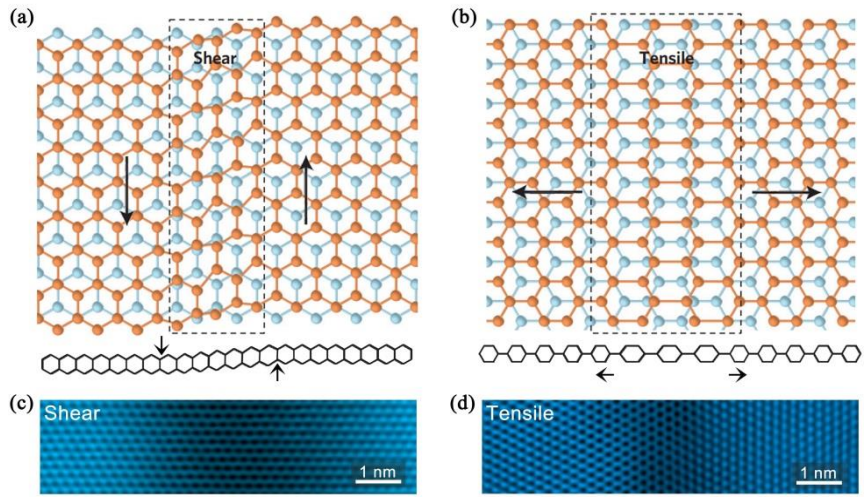

**Fig. 15** Shear and tensile stacking domain wall solitons. **(a, b)** Schematics of the shear- and tensile-type strain solitons, respectively. **(c, d)** Atomic STM images of the shear and tensile domain walls, respectively. The arrows indicate the displacement vectors. Reproduced from Refs. [150] and [151].

### 4.1.1 AB-BA bilayer soliton

An AB-BA bilayer domain wall ~8 nm in width on graphite surface were studied [Fig. 16(a)]. In STM measurements, the electronically decoupled bilayer graphene on graphite exhibited small-period Moiré patterns because of the large rotation angle with respect to the substrate [73]. The ultra-low random potential fluctuations resulting from substrate imperfections allows us to obtain high-quality atomic-resolution STM images of the 1D domain wall, as shown in Fig. 16. Figure 16(b) shows a representative atomic STM image of the AB-BA stacking soliton. From the left to the right of the domain wall, the atomic image transforms from a triangular lattice (in the AB region) to a hexangular lattice (in the center of the domain wall) and then to a triangular lattice (in the BA region) [157, 158], completing an interlayer transition from AB to BA stacking. Moreover, the interatomic distances in the domain wall are ~1.5 % smaller than those in the surrounding Bernal regions, according to the 2D Fourier transform of the atomic STM images. The compressed interatomic distances of the 1D structure, together with the hexagonal lattice at its center and the surrounding AB and BA domains, imply that the structure is an AB-BA strain soliton. As shown below, to identify the AB-BA stacking domain wall, LL spectroscopic measurements are necessary.

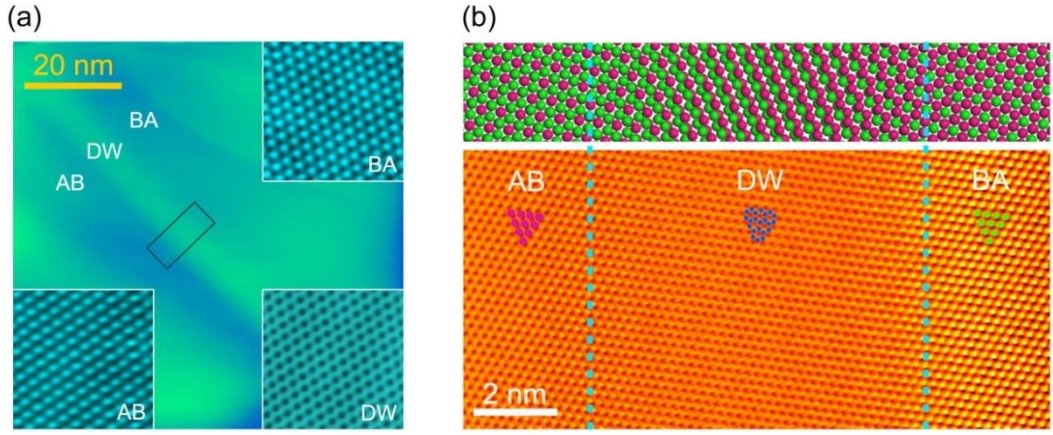

**Fig. 16** AB-BA bilayer domain wall imaged by STM. **(a)** STM topographic image of an AB-BA domain wall (DW) region on a graphite surface. Insets: atomic-resolution STM images in the AB, DW, and BA regions. **(b)** Lattice transition from a triangular lattice (in the AB region) to a hexangular lattice (in the center of the domain wall) and then to a triangular lattice (in the BA region). Reproduced from Ref. [73].

The STS spectra recorded in both the AB- and BA-stacked regions exhibit characteristics that are expected for gapped graphene bilayers (see Fig. 17). The graphite substrate induces an interlayer bias and breaks the inverse symmetry of the topmost adjacent bilayers, generating a finite gap of ~80 meV in the parabolic bands of both the AB and BA bilayers. At the level of low-energy effective theory, the AB-stacked bilayer is equivalent to the BA-stacked bilayer subjected to the opposite gate polarity [159–161]. Thus, under a uniform interlayer bias, the sign of the energy gap (as well as the effective mass) changes across the domain wall from the AB- to BA-stacked regions, and symmetry-protected gapless modes are expected to emerge in the stacking soliton [152, 153]. Under finite magnetic fields, the positions of the two lowest LLs—$LL_{(0,1,+)}$ and $LL_{(0,1,-)}$, which are a couple of layer-polarized quartets—in the energy band depend on the sign of the gate polarity (or the sign of the energy gap) of the gapped Bernal bilayers [45, 72, 77]. Therefore, they are reversed in the AB- and BA-stacked bilayers. In our experiment, the LL spectra recorded in the AB- and BA-stacked bilayer domains exhibited strong Landau quantization of massive Dirac fermions of gapped bilayers, as shown in Figs. 17(a) and 17(e). We obtained the same effective mass—$m^* \approx 0.045\, m_e$—in the two domains. The most remarkable feature is that the top layer-located quartet $LL_{(0,1,+)}$ lies on the valence-band side for the AB-stacked bilayer region but on the conduction-band side for the BA-stacked region. This result directly demonstrates that the band structures of the AB and BA bilayers are reversed, leading to the change in the sign of the energy gap. This feature in the LL spectra can be used to unambiguously identify the AB-BA domain-wall region in bilayer graphene.

Next, we discuss the Landau quantization in the AB-BA stacking soliton region. The bilayer stacking domain walls are predicted to generate topological conducting edge states in the gated system, which was proven by STM [73] and transport measurements [155]. Under strong magnetic fields, we can also detect a series of DOS peaks in the stacking domain wall, as shown in Figs. 18 and 19. These peaks mimic the sequences of the LLs of both the AB and BA domains. This feature results from the spatial extension of the LL wave functions of the surrounding Bernal bilayers (the extent, $\sim 2\sqrt{N}l_B$, is on the order of 10 nm for $B \leq$

8 and comparable to the soliton width), as described in Section 3, which leads to the appearance of the bilayer-like LLs in the domain-wall region. The "splitting" of the LLs recorded in the domain wall, as shown in Figs. 18(a) and 18(b), arises from the shift of the CNPs of the adjacent AB and BA domains (~15 meV in the experiment). The calculated LLs in the domain wall under a 8-T field imitate the sequences of those in the AB and BA regions, which agrees well with the experimental results [see Fig. 18(c)]. Our results indicate that the layer-stacking domain walls, as well as the stacking orders, can be unambiguously identified by LL spectroscopic measurements.

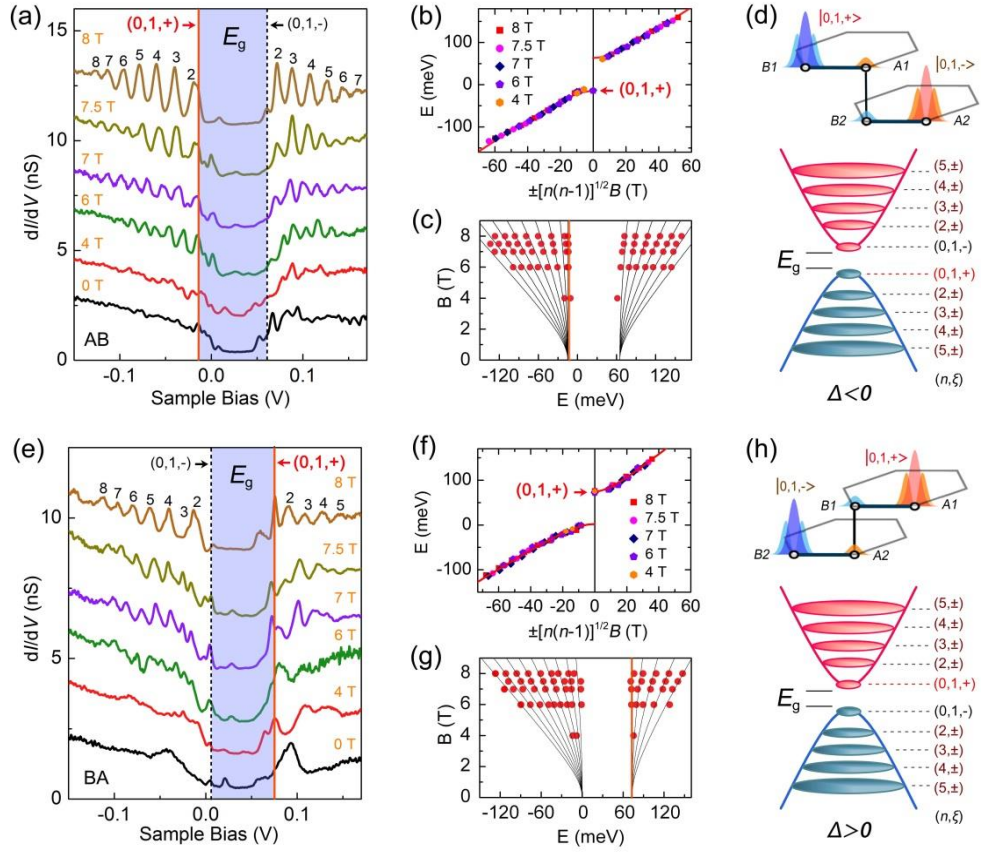

**Fig. 17** Landau quantization of gapped AB and BA bilayer graphene. **(a, e)** LL spectra recorded in the AB- and BA-stacked regions under various fields $B$, respectively. **(b, c)** LL peak energies of the AB bilayer plotted versus $\pm[n(n-1)]^{1/2}B$ and $B$, respectively. The solid curves are the fits of the data with the theoretical equation. **(d, h)** Schematics of the layer-polarized states and the LLs in gapped AB and BA bilayer graphene, respectively. **(f, g)** LL peak energies of the BA bilayer plotted against $\pm[n(n-1)]^{1/2}B$ and $B$, respectively. Reproduced from Ref. [73].

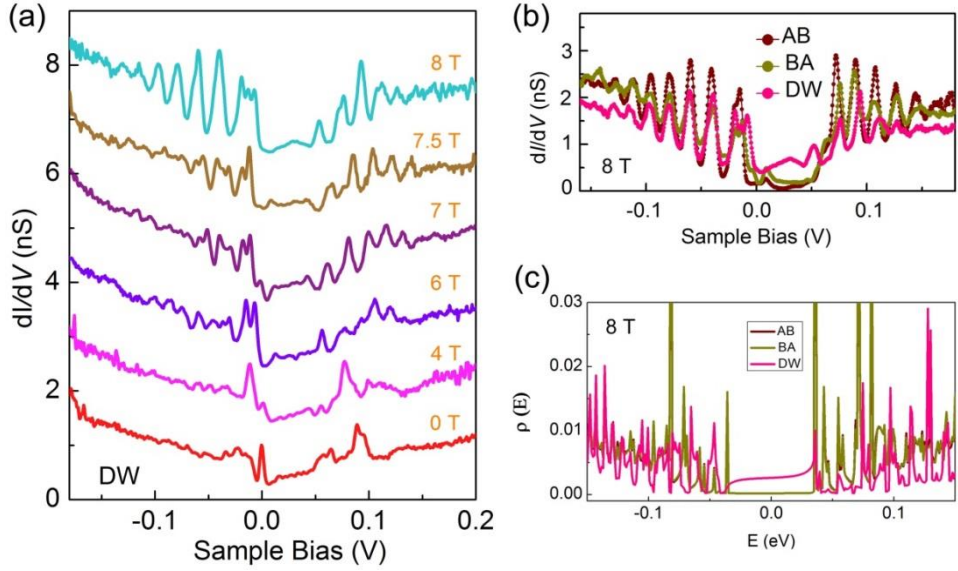

**Fig. 18** STS spectra of the bilayer domain wall under various fields $B$. **(a)** STS spectra of the domain-wall region for different fields $B$, showing level peaks. **(b)** LL spectra of the AB, BA, and domain-wall regions under an 8-T field. **(c)** Calculated DOS of the AB, BA, and domain-wall regions under an 8-T field. Reproduced from Ref. [73].

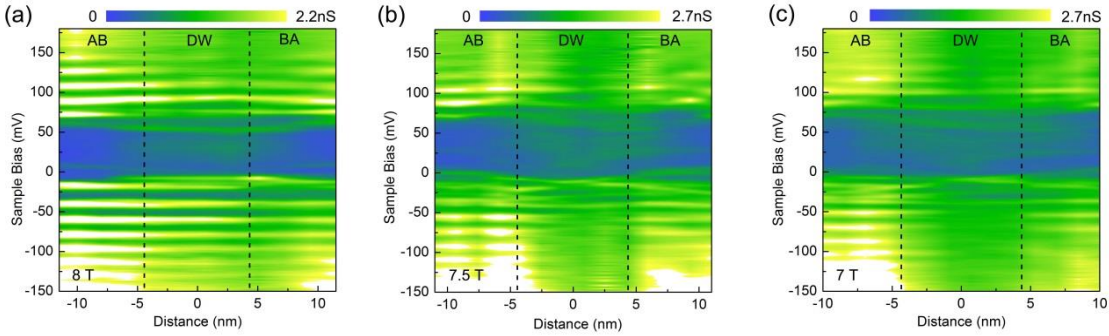

**Fig. 19** Spatially resolved tunneling spectra maps across the AB-BA domain wall. The maps are measured at 8 T **(a)**, 7.5 T **(b)**, and 7 T **(c)**. The LLs shown in the domain-wall region almost follow the sequences of those recorded in the gapped AB/BA bilayer region. The dashed lines indicate the edges of the domain wall. Reproduced from Ref. [73].

### 4.1.2 ABA-ABC trilayer soliton

In trilayer graphene, there are two lowest-energy stacking orders: ABA, i.e., Bernal stacking, and ABC, i.e., rhombohedral stacking [Figs. 20(a) and 20(b)]. The chiral quasiparticles differ significantly between the two allotropes because of the distinction in the stacking order [162–164]. In the ABA trilayer, both massless ($l = 1$) and massive ($l = 2$) Dirac fermions coexist [89, 95]; whereas the low-energy excitations in the ABC trilayer are $l = 3$ chiral quasiparticles with a cubic dispersion (the corresponding Berry phase of the quasiparticles is $3\pi$) [165–167]. Moreover, stacking domain-wall solitons separating the ABA and ABC registered regions can be formed in trilayer graphene. By using similar STM measurements as described for bilayers, ABA- and ABC-stacked trilayers, together with the ABA-ABC trilayer domain walls, were observed in our experiment [150].

Figures 20(c)–20(f) show several STM images taken around the stacking solitons of trilayer graphene. The ABA region can be directly discriminated from the ABC region in the STM images recorded at low bias voltages owing to its distinct low-energy electronic structures and properties, as shown in Figs. 20(a) and 20(b). The bright lines separating the adjacent ABA and ABC regions in the topography images are the trilayer stacking domain-wall solitons, whose heights depend strongly on the bias voltage used for imaging. Furthermore, the topographical structures of the trilayer domain walls exhibit different patterns, as shown in Figs. 20(c)–20(f). These rich patterns provide an ideal platform for exploring stacking solitons with different atomic configurations. The representative atomic-resolved STM image around a trilayer domain wall shows hexagonal lattices in the center of the solitons, but triangular lattices are observed in both the adjacent ABA and ABC regions [see the lower inset of Fig. 20(f)]. The types of the trilayer domain walls are determined according to the lattice transition across the solitons. In the experiment, we observed shear-type domain walls more frequently than in the case of the tensile-type solitons in the trilayers. Similar results were reported for the stacking domain walls of bilayer graphene [147, 151], which may arise from the fact that the energy of the shear stacking solitons is slightly lower than that of the tensile stacking solitons in multilayer graphene.

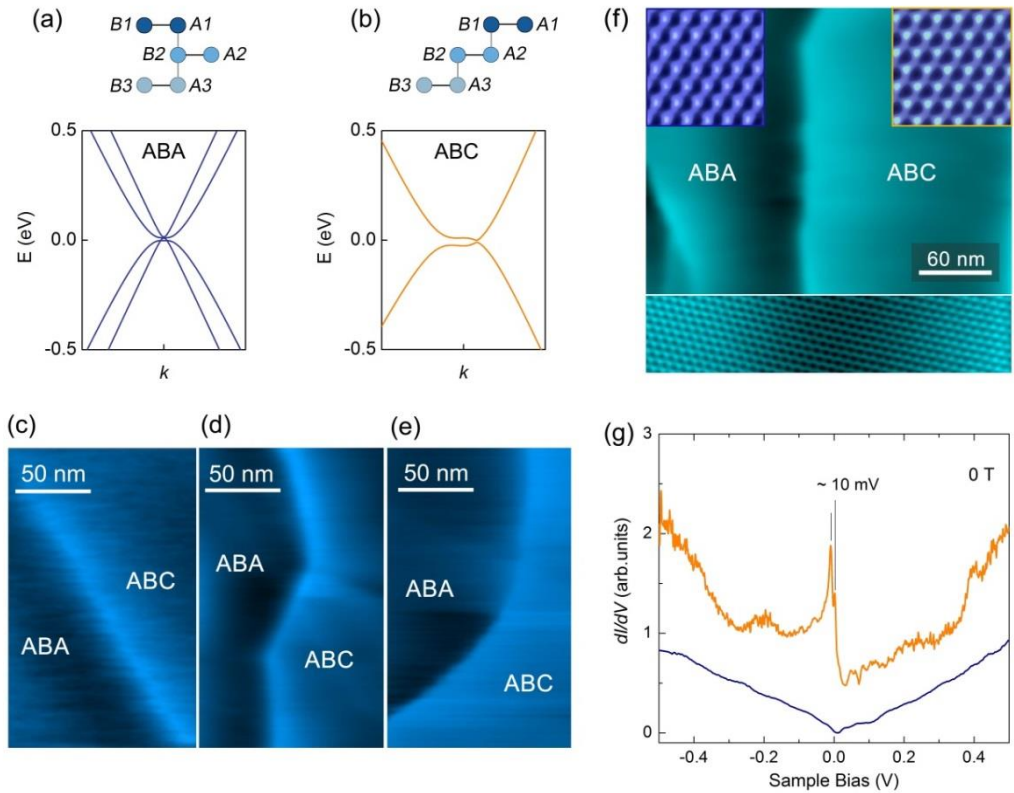

**Fig. 20** Schematics and band dispersions of ABA **(a)** and ABC **(b)** trilayer graphene. **(c–f)** STM topographic images of different ABA-ABC trilayer domain-wall structures observed on the graphite surface. Insets in (f): atomic lattices in the ABA, ABC, and domain-wall regions. **(g)** Typical zero-field tunneling spectra of the ABA and ABC trilayers. Reproduced from Ref. [150].

Figure 20(e) shows the STS spectra recorded in the ABA and the ABC regions away from the stacking solitons, under zero magnetic field. For the ABA trilayer, we observed a typical *V*-shaped spectrum at 0 T, as previously reported. For the ABC trilayer, the zero-field spectrum exhibited low-energy pronounced peaks, which were generated by the flat bands around the CNP of the ABC trilayer graphene [168]. The energy spacing ~10 meV of the two peaks around the CNP corresponds to the energy gap of the ABC trilayer. The substrate induces an effective interlayer bias, breaking the inversion symmetry of the rhombohedral trilayer [169] and thus introducing a gap, which is similar to that in bilayer graphene.

The finite-field spectra (Fig. 21) exhibit quite different Landau quantization in ABA-stacked and ABC-stacked graphene trilayers. In the STS spectra of the ABA trilayer, we observed a sequence of LL peaks of both massless Dirac fermions ($l = 1$) and massive Dirac fermions ($l = 2$), as discussed in Section 2. For the ABC-stacked trilayer, the tunneling spectra show a new LL sequence of the unique Landau quantization for the $l = 3$ chiral quasiparticles [Fig. 21(d)]. In standard theory, the $l = 3$ chiral fermions of ABC trilayers are quantized under a perpendicular magnetic field with the energy of the $n^{th}$ level, which has a $B^{3/2}$ dependency [91, 170, 171]:

$$E_n = E_C \pm \frac{2\hbar v_F^2 eB}{t_\perp^2}^{3/2} \sqrt{n(n-1)(n-2)}, \qquad n = 3, 4, 5..., \qquad (7)$$

where $E_C$ is the energy of the CNP, ± describes electrons and holes, and $t_\perp$ is the nearest-neighbor interlayer hopping strength. By separately fitting the energies of the LLs in the electron and hole branches to Eq. (7), as shown in Figs. 21(e) and 21(f), we obtained a bandgap of $Eg \approx 10$ meV for the ABC trilayer, which is congruent with that observed under zero field. We also determined the interlayer coupling to be $t_\perp \approx 0.48$ eV in different ABC samples. It was theoretically predicted that $t_\perp \approx 0.5$ eV in the ABC trilayer graphene [165], and an almost identical value of $t_\perp$ was extracted from transport measurements in the ABC trilayers [162].

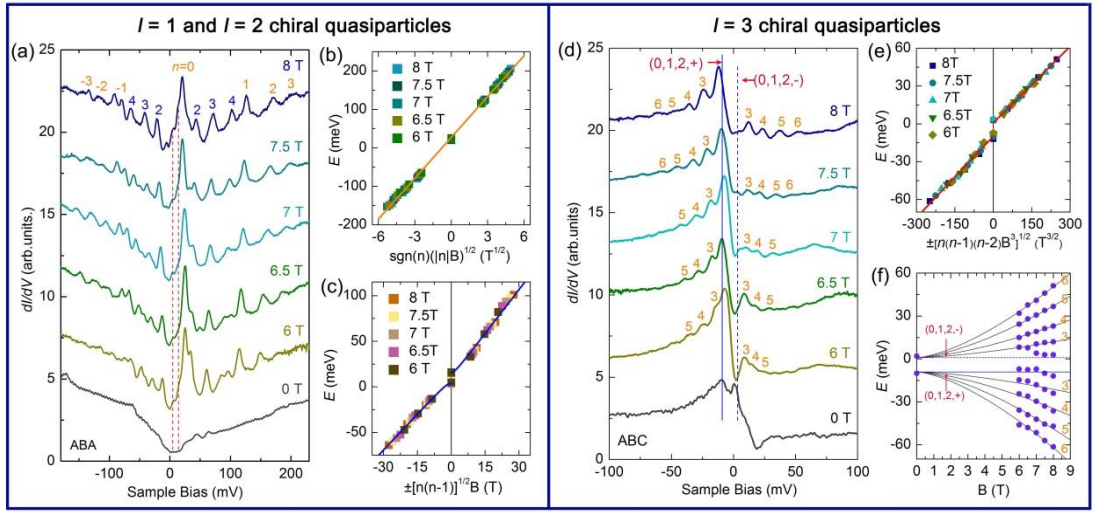

**Fig. 21** (a) STS spectra of the ABA trilayer measured under various fields $B$. The monolayer and bilayer LL orbital indices are indicated by orange and blue numbers, respectively. The dashed lines label the gap edges in the parabolic bands. LL peak energies obtained from (a), showing **(b)** $B^{1/2}$ and **(c)** $B$ dependency, respectively. **(d)** Tunneling spectra of the ABC trilayer under 0–8-T magnetic fields. The LL indices are labeled. **(e)** LL peak energies obtained from (d) versus $\pm[n(n-1)(n-2)B^3]^{1/2}$. **(f)** LL peak energies versus $B$. The blue dots are the data from (g), and the black curves are results of fitting with Eq. (7). Reproduced from Ref. [150].

The LL is also fourfold degenerate in ABC-stacked trilayer graphene, with two valley degeneracies and two spin degeneracies for each orbital quantum number, $n$. Without the roles of an external electric field and

electron–electron interaction, the $n = 0$, $n = 1$, and $n = 2$ LLs are further degenerate and consequently, there is a 12-fold degenerate state at the CNP of the ABC trilayers. The degeneracy of the lowest LL can easily be partially lifted. For example, in the presence of an interlayer potential, the valley degeneracy is lifted and a finite energy gap is generated in the band structure of ABC trilayer graphene [172]. Moreover, the wave functions of the lowest LL are mainly localized on the *A1* sites of the first layer for one valley (+) and on the *B3* sites of the third layer for the other valley (-) [173]. Hence, we observed valley-polarized Landau quantization of the $l = 3$ chiral quasiparticles: the magnitude of the valley-polarized LL—$LL_{(0,1,2,+)}$—was far higher than that of the $LL_{(0,1,2,-)}$ because the STM predominantly probes the DOS of the top graphene layer, as shown in Fig. 21(d). The layer polarization of the two lowest LLs, $LL_{(0,1,2,+)}$ and $LL_{(0,1,2,-)}$, depends on the sign of electric polarity (or the sign of the energy gap) of the ABC trilayer, and we observed both positive and negative layer-polarized $LL_{(0,1,2,+)}$ and $LL_{(0,1,2,-)}$ in different trilayers [150]. Theoretically, topological edge modes are expected to be observed in such trilayer domain walls [154], which separates opposite gated ABC-stacked domains, as demonstrated in the AB-BA domain walls of bilayer graphene [73].

Now, we discuss the Landau quantization of the quasiparticles across the stacking solitons between the ABA and ABC trilayer domains. Figure 22 shows representative measured LL spectra with respect to the STM tip position, scanned from the ABA region to the ABC region. Away from the trilayer domain wall, we observed LLs of the $l = 3$ chiral fermions in the ABC trilayer and detected LLs of both the $l = 1$ and $l = 2$ chiral quasiparticles in the ABA trilayer. Across the stacking soliton, we observed the evolution of quasiparticles between the chiral degree $l = 1\&2$ in the ABA trilayer and $l = 3$ in the ABC trilayer, accompanying the transition of the stacking orders of the trilayers. In the stacking domain-wall region, the LLs of both the $l = 1\&2$ and $l = 3$ chiral fermions are detected because of the spatial extension of the wave functions for the quasiparticles in the adjacent domains [72].

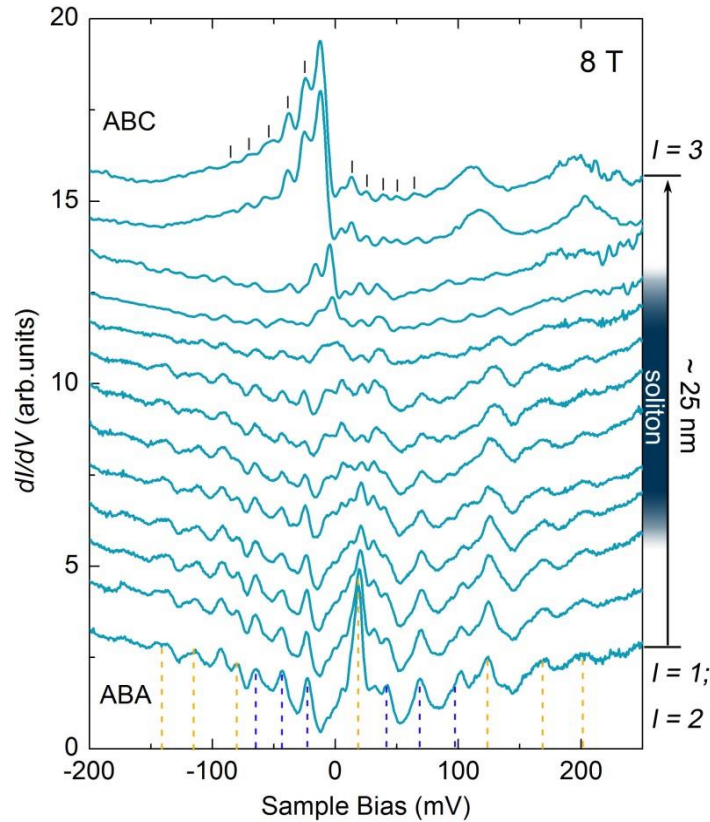

**Fig. 22** Spatial evolution of the LL spectra recorded under an 8-T field across an ABA-ABC shear soliton. The colored dashed lines indicate the positions of the LLs of the massless and massive fermions in the ABA trilayer. The black bars label the LLs of the quasiparticles in the ABC trilayer. Reproduced from Ref. [150].

### 4.2 Twisted graphene system

In addition to the matched layer-configurations, the introduction of a stacking misorientation drastically expands the allotropes of the graphene multilayers and, more importantly, the resultant structures show strong twist-dependent electronic spectra and properties [105, 174–177]. For example, two low-energy van Hove singularities (VHSs), which originate from the two saddle points in the band structure, were observed in the twisted bilayer graphene as two pronounced peaks in the DOS [178–181]. Among such especial dispersion of the twisted bilayers, the most striking results are the angle- and coupling-dependent renormalization of the Fermi velocity [182, 183] and the appearance of almost dispersionless bands (flat bands) at a very small angle [184]. This suggests that non-Abelian gauge potentials emerge in the smallest twisted bilayers (≤1°) [185] and that electrons in a graphene bilayer can be changed from ballistic to localized by simply varying the rotation angle [186–188]. Here, we discuss the angle-dependent Landau quantization in twisted graphene bilayers and trilayers.

#### 4.2.1 Twisted bilayers

In the continuum approximation, a rotation between two graphene layers leads to a shift $\Delta K$ [Fig. 23(c)] between the corresponding Dirac points of the two sheets in the momentum space [189, 190]. The zero-energy states no longer occur at $k = 0$ but rather occur at $k = -\Delta K/2$ for layer one and $k = \Delta K/2$ for layer two [191, 192]. At a large rotation angle ($\theta > 6°$), the twisted bilayers behave as two decoupled single-layer graphene sheets with a Fermi velocity of $v_F \approx 1.1 \times 10^6$ m/s [181, 182]. At a small rotation angle ($\theta < 6°$), the electronic states near the Dirac cones of two layers couple with a reduced interlayer hopping amplitude of $t_\theta \approx 0.4 t_\perp$ ($t_\perp$ is the interlayer hopping for *AB*-stacked layers) [179, 182, 193]. As a result, the two Dirac cones intersect and hybridize at the energy of $\pm \hbar v_F \Delta K$, generating two saddle points in the energy bands [Fig. 23(d)] and two symmetric low-energy VHSs in the DOS. The energy difference between two the VHSs exhibits a strong angle-dependent variation: $\Delta E_{VHS} \approx \hbar v_F \Delta K - 2 t_\theta$ [Fig. 23(e)]. Below the energy of the VHSs, the linear dispersion is preserved but with a renormalized Fermi velocity.

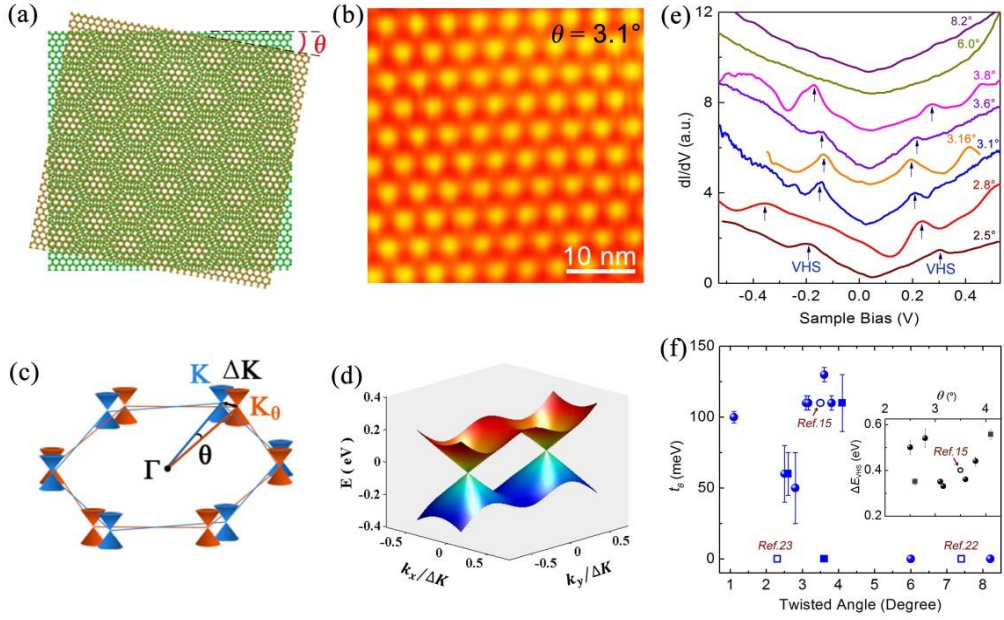

**Fig. 23** Small-angle twisted bilayer graphene. **(a)** Schematic of the twisted bilayer. **(b)** STM topographic image of a twisted bilayer with $\theta = 3.1°$, showing the Moiré superlattice structure. **(c)** Shifting of the Dirac cones in twisted bilayers. **(d)** Low-energy band dispersion of the twisted bilayer with $\theta = 2.2°$ and $t_\theta = 156$ meV. **(e)** STS spectra of twisted graphene bilayers with various twisted angles. The arrows mark the positions of the VHSs. **(f)** Interlayer hopping and energy differences of the VHSs (the inset) with respect to the rotation angles in twisted graphene bilayers. Reproduced from Refs. [182] and [196].

The Fermi velocity of a twisted bilayer can be obtained by LL spectroscopic measurements [182]. In the twisted bilayers, two series of LL sequences are generated below the VHSs under strong magnetic fields, which arise from the two shifted Dirac cones of the two layers. Each of the LL sequences shows the same square-root dependency of both the level index $n$ and field $B$ as in single-layer graphene, i.e., $E_n = E_D + \mathrm{sgn}(n)\sqrt{2e\hbar v_F^2 |n| B}$, $n = 0, \pm1, \pm2,\ldots$ ($E_D$ is the energy of the Dirac point). Figure 24 shows several representative Landau quantized spectra and their analysis for different twisted bilayer samples. The sequence of LLs shown in Figs. 24(a)–24(c) is unique to that of massless Dirac fermions and is expected to be observed in twisted bilayer graphene. In the LL spectroscopic measurements, we only measured one LL sequence, as the quasiparticles of the topmost graphene layer mainly come from one of the two Dirac cones in the twisted bilayers and the STS predominantly probes the signal of the top layer. The Fermi velocities of the twisted bilayers with various rotation angles can then be obtained directly by a reasonable linear fit of the measured LL energies to the above equation, as shown in Figs. 24(d)–24(f).

Our experimental results indicate that both the twisted angles and the interlayer-coupling strengths drastically affect the Fermi velocity of the twisted bilayers. In the twisted graphene bilayers, the stacking fault, grain boundary, defects, and roughness of the substrate may alter the interlayer distance and stabilize it at various equilibrium values, leading to large variations of the interlayer interaction [182]. For a constant $t_\theta$, the Fermi velocity decreases as the twisted angle decreases. This tendency can be unambiguously displayed in the band structures of different twisted bilayers with identical interlayer coupling, as shown in Fig. 25(a). The slope of the energy dispersion near the Dirac point, which is proportional to the Fermi velocity, decreases as the rotation angle decreases. Normally, the interlayer-coupling strength $t_\theta \approx 110$ meV in the small twisted bilayer graphene, which can be obtained from the relationship between $t_\theta$ and $\Delta E_{VHS}$ [179, 181, 182]. In this case, the renormalization of the Fermi velocity in twisted bilayers can be described as follows [189]:

$$\frac{\tilde{v}_F}{v_F} = 1 - 9\left(\frac{t_\theta}{\hbar v_F \Delta K}\right)^2. \qquad (8)$$

Hence, the pronounced Fermi velocity renormalization is observed in small twisted bilayers, as shown in Fig. 24(b), where the measured Fermi velocity is $v_F \approx 0.81 \times 10^6$ m/s for a twisted bilayer with $\theta \approx 3.6°$ and $t_\theta \approx 130$ meV. However, for relatively weak interlayer hopping, the renormalization of the Fermi velocity is negligible, and we cannot detect the significant reduction of the Fermi velocity, even for small-angle twisted bilayers. For example, in the $\theta \approx 2.8°$ twisted bilayer with interlayer hopping, $t_\theta \approx 50$ meV. The weak interlayer hopping may arise from the tilt grain boundary underlying the twisted bilayers. The Fermi velocity is measured to be $v_F \approx 1.03 \times 10^6$ m/s, showing a negligible renormalization.

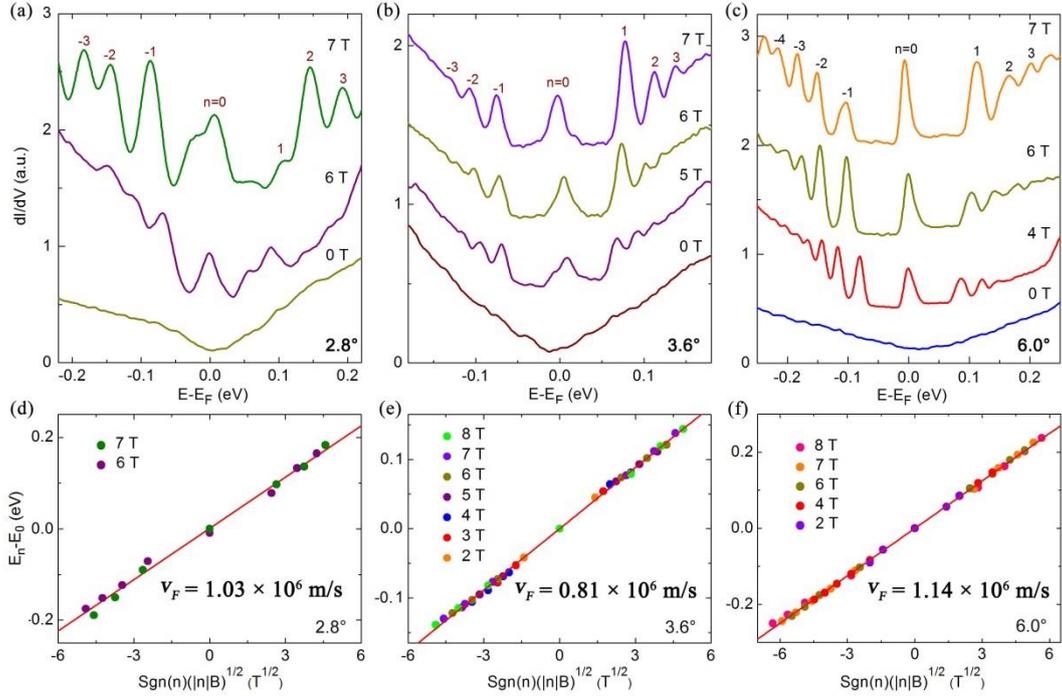

**Fig. 24** Landau quantization in twisted bilayer graphene. STS spectra taken under various field $B$ in twisted bilayers with **(a)** $\theta = 2.8°$, **(b)** $\theta = 3.6°$(b), and **(c)** $\theta = 6.0°$. The LL indices are marked. **(d–f)** LL peak energies obtained in (a)–(c), respectively, plotted against sgn(n)(|n|B)$^{1/2}$, as expected for massless Dirac fermions. The solid lines are linear fits of the data. The slopes yielding the Fermi velocities of $v_F = (1.03 \pm 0.02) \times 10^6$ m/s (d), $v_F = (0.811 \pm 0.004) \times 10^6$ m/s (e), and $v_F = (1.140 \pm 0.003) \times 10^6$ m/s (f), respectively. Reproduced from Ref. [182].

The effects of both the interlayer-coupling strength and the twisted angle on the Fermi velocity of the twisted bilayers are shown in Fig. 25(b). The observed Fermi velocity does not decrease monotonously as the twisted angle decreases, indicating that it is not determined only by the twisted angle but also depends on the interlayer interaction. By using the discrete interlayer-coupling strength, the Fermi velocity shows the predicted angle dependence of the renormalization for different twisted angles, as described by Eq. (8).

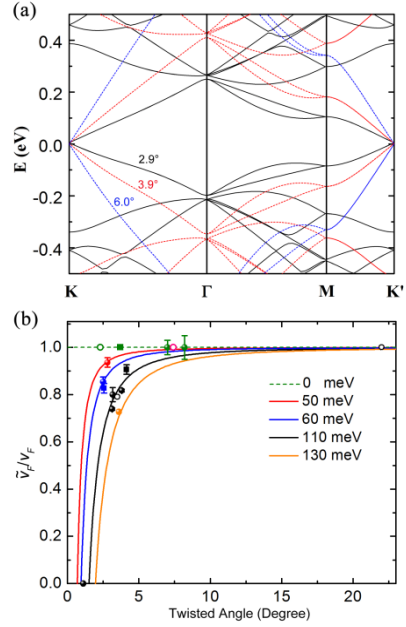

**Fig. 25 (a)** Energy dispersions for three twisted graphene bilayers with $\theta = 6.0°$, $\theta = 3.9°$, and $\theta = 2.9°$, respectively. **(b)** The Fermi velocity as a function of the twisted angle for different interlayer hopping strengths. The circles and squares indicate the data obtained for HOPG and SiC, respectively. The Fermi velocity is normalized with respect to $v_F = 1.1 \times 10^6$ m/s. Reproduced from Ref. [182].

### 4.2.2 Twisted trilayers

We discuss twisted trilayers consisting of two Bernal-stacked graphene bilayers and one slightly rotated layer [Fig. 26(a)]. The low-energy band structures of a twisted trilayer are formed by a parabolic band resulting from the *AB*-stacked bilayer intersecting with a Dirac cone band originating from the rotated layer [Fig. 26(b)] [194, 195]. Similar to the case in twisted bilayers, the two band branches of the twisted trilayers hybridize in a finite interlayer coupling between the rotated layer and the Bernal bilayer. Hence, two saddle points are generated in the energy dispersion, and two VHSs are generated in the DOS [Figs. 26(c) and 26(d)]. However, with an identical twisted angle and identical interlayer coupling, the energy difference of the saddle points in the electronic band structures of the twisted trilayers, i.e., $\Delta E_{VHS}$, is far smaller than that of the twisted bilayers [196, 197].

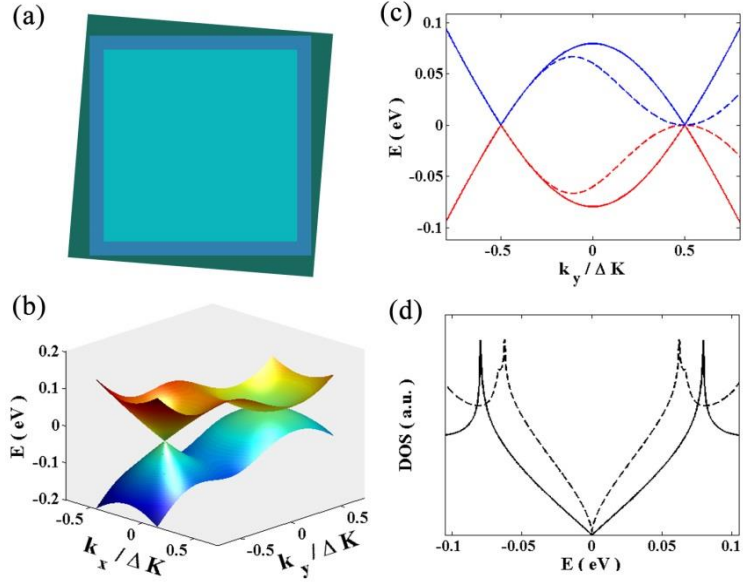

**Fig. 26** Configuration and band structures of twisted trilayer graphene. **(a)** Schematic of the twisted trilayer graphene made out of two Bernal bilayers and one rotated layer. **(b)** 3D low-energy band structure of a twisted trilayer with $\theta = 2.2°$ and $t_\theta = 156$ meV. **(c)** Band structures along $k_x = 0$ for the twisted trilayer (dashed curve) and bilayer (solid curve) with the same $\theta$. **(d)** Calculated DOS of the twisted trilayer (dashed curve) and bilayer (solid curve) with the same $\theta$, both showing two VHSs. Reproduced from Ref. [196].

Figure 27 shows the coexistence of the twisted graphene trilayer and the twisted bilayer regions, which are separated by a tilt grain boundary and both have a twisted angle of 2.8° [197]. The tilt grain boundary is located at the second graphene layer, and the top continuous layer is rotated with respect to the second layer, resulting in the twisted bilayer in region I, as schematically shown in Fig. 27(d). In region II, the top graphene layer and second layer are AB-stacked, and the Moiré pattern arises from the stacking misorientation between the second layer and third layer [196]. Consequently, the contrast of the Moiré pattern for the AB-twisted trilayer (ABT, region II) is much lower than that for the twisted bilayer (TB, region I), as shown in the STM images of Figs. 27(a)–27(c). Figure 28(a) shows the spatial evolution of the zero-field tunneling spectra along a line across the tilt grain boundary from TB to ABT. In both the TB and ABT regions, the spectra exhibit two low-energy VHSs. However, the energy difference of the VHSs—$\Delta E_{VHS}$—abruptly decreases from ~0.54 eV in TB to ~0.44 eV in ABT, as shown in Fig. 28(c). This decrease in $\Delta E_{VHS}$ is a direct result of introducing a third layer on top of a twisted bilayer, as previously discussed. In the ABT, the substrate induces an effective external electric field on the topmost *AB*-stacked bilayer, which consequently generates a finite gap ~100 meV in the parabolic band branch of the ABT. The spectra recorded in the ABT region at positions far from the boundary (>2.4 nm) display two DOS peaks at the gap edges within the two VHSs.

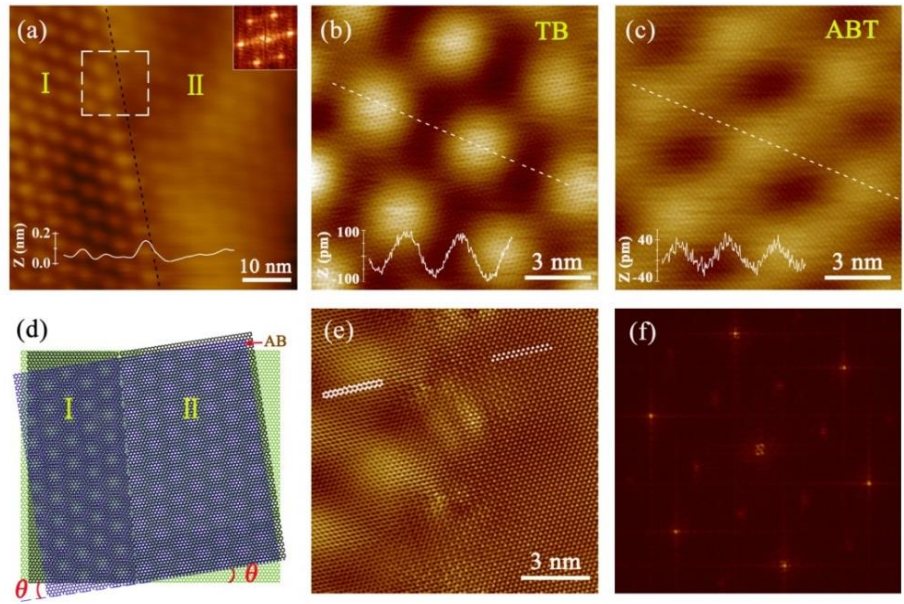

**Fig. 27** Coexistence of twisted graphene trilayer and bilayer. **(a)** STM images of twisted graphene trilayer (region I) and bilayer (region II) on a graphite surface. The two sets of Moiré superlattices are separated by a tilt grain boundary (black dashed line). The periods of both the Moiré patterns ~5.0 nm. Inset (bottom): a profile line across the boundary. Inset (top): Fourier transforms of the STM image. STM images of a twisted **(b)** bilayer and **(c)** trilayer. Insets: height profiles along the white dashed lines. **(d)** Schematic of the structure in (a). **(e)** High-resolution current image of the area indicated by the white frame in (a). The topmost graphene layer is continuous. It shows a clear hexagonal lattice in region I and a triangular lattice in region II. **(f)** Fourier transforms of (e). The outer six bright spots represent the reciprocal lattice of the topmost graphene layer. Reproduced from Ref. [197].

Under strong magnetic fields, we observed distinct Landau quantization in the TB and ABT. In the TB region, the STS spectra exhibit the expected one LL sequence of the massless Dirac Fermion for the twisted graphene bilayer [197]. In the ABT region, the observed LL spectra [Fig. 28(d)] are distinct from those in the TB region and follow those of a massive Dirac fermion [72]. Such a result is reasonable because the massive Dirac fermions of the ABT are localized at the topmost *AB*-stacked bilayer. For the massive Dirac fermions, two DOS peaks at the gap edges are expected to develop into two valley-polarized fourfold-degenerate LLs in finite fields [77]. This is demonstrated explicitly in our experiment, as shown in Fig. 28(d). The positions of the two lowest LLs are almost independent of the magnetic field, meaning that the gap size in the parabolic band of the twisted graphene trilayer does not vary with the field *B*. We expect that if the rotated sheet is the topmost one in the twisted graphene trilayer, the measured LL sequence will follow that of massless Dirac fermion in the STS probing.

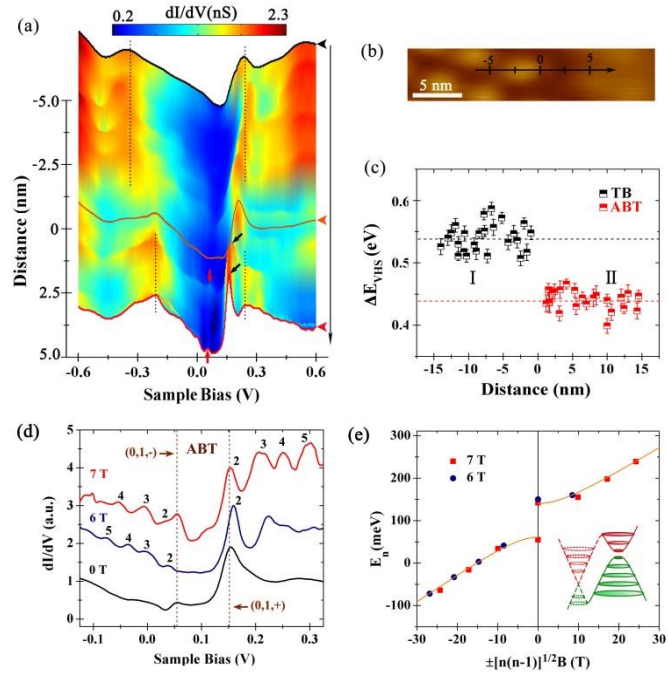

**Fig. 28 (a)** STS spectroscopy as a function of the tip position measured along the black arrow in (b). The black dotted lines indicate the positions of VHSs in the TB and ABT regions. Two peaks (labeled by black and red arrows) in the spectra of the ABT region mark the positions of the DOS peaks at the gap edges. **(b)** An STM image obtained around the boundary of the TB and ABT regions. **(c)** $\Delta E_{VHS}$ as a function of the position around the boundary. The black and red dashed lines show the average values for the TB (~0.54 eV) and ABT (~0.44 eV) regions, respectively. **(d)** $dI/dV$ spectra measured in the ABT region for different fields $B$. LL indices $n$ are marked. The zero-energy LL splits into two valley-polarized quartets and a bandgap of ~100 meV is opened, as indicated by the dashed lines. The $LL_{(0,1,+)}$ and $LL_{(0,1,-)}$ are projected on the top and bottom graphene layers, respectively. **(e)** LL peak energies shown in (d) plotted against $\pm[n(n-1)]^{1/2}B$. The red lines are fits of the data. The inset shows a schematic of the low-energy dispersion of ABT with quantized LLs. Reproduced from Ref. [197].

## 5   Landau quantization in strained and defective graphene

Being only one layer thick, the structure of graphene is vulnerable to the underlying substrate and surrounding environment [4]. Consequently, there are many distortions, such as corrugations and charge defects, in supported graphene, especially on metal substrates [198–200]. Around these distortions, the properties of quasiparticles are drastically modified, leading to unusual spectroscopic characteristics [201, 202]. For example, both the strained structures and charge defects will introduce an electron–hole asymmetry in graphene, and a spatially varying lattice distortion in strained graphene can create pseudo-magnetic fields, resulting in LL-like quantization [203, 204]. In this section, giant electron–hole asymmetry and unconventional valley-polarized LL splitting are investigated by revealing the Landau quantization in strained and defective graphene under external magnetic fields.

## 5.1 Electron–hole asymmetry

Charge carriers in a graphene monolayer exhibit light-like dispersion and, usually, the electron and hole are symmetric [4, 5]. The electron–hole symmetry plays a crucial role in the chirality and chiral tunneling of massless Dirac fermions in graphene monolayers [6]. Previous researches demonstrate that strain and charged defects can induce large electron–hole asymmetry in graphene [205–208]. According to the Landau quantization under a magnetic field, the Fermi velocities of electron- and hole-like Dirac fermions in a graphene monolayer can be determined separately. This provides a unique opportunity for quantitatively studying the electron–hole asymmetry. Next, via LL spectroscopy, we show the large electron–hole asymmetry generated by strain and charged-defect scattering in graphene monolayers [76].

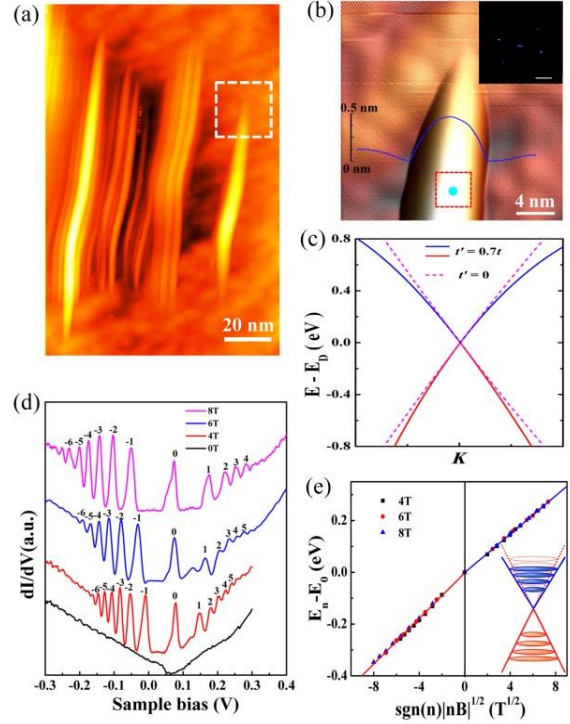

**Fig. 29** Electron–hole asymmetry in strained graphene. **(a)** An STM topographic image showing quasi-periodic graphene ripples on Rh foil. **(b)** STM image of a graphene ripple in the white frame in (a). The solid blue line is the height profile across the rippled graphene region. Inset: fast Fourier transform showing the reciprocal lattice of the graphene lattice. The scale bar represents 20 Gm$^{-1}$. **(c)** Electronic dispersion for a graphene monolayer with different $t'$. **(d)** STS spectra taken on the ripple in (b) with different fields $B$. **(e)** LL peak energies plotted against sgn($n$)$|n|B^{1/2}$. The red and blue solid lines are linear fits of the data with Eq. (1) for electrons and holes, respectively. Inset: the schematic LLs structure of graphene with (solid curves) and without (dashed curves) strain. Reproduced from Ref. [76].

The Hamiltonian of the graphene monolayer in the tight-binding model is [5]

$$H = -t\sum_{\langle i,j \rangle}(a_i^\dagger b_j + \text{H.C.}) - t'\sum_{\langle\langle i,j \rangle\rangle}\left(a_i^\dagger a_j + b_i^\dagger b_j + \text{H.C.}\right) \quad (9)$$

Here, the operators $a_i^\dagger$ ($a_i$) create (annihilate) an electron with spin at site $i$, $t \sim 3$ eV is the nearest-neighbor hopping integral, and $t'$ is the next-nearest-neighbor hopping energy. According to Eq. (9), the simplest method to break the electron–hole symmetry in a graphene monolayer is to introduce a nonzero $t'$, which is easily done in strained graphene [199, 205]. Figures 29(a) and 29(b) show representative STM images of a rippled graphene region on Rh foil due to the mismatch of thermal-expansion coefficients between the graphene and the Rh substrate. STS spectra recorded in the ripples under magnetic fields exhibit the Landau quantization of massless Dirac fermions, as shown in Fig. 29(d). According to Eq. (1), we determined $v_e^F$ and $v_h^F$ separately. The Fermi velocities of electrons, $v_e^F$, and holes, $v_h^F$, differ significantly, as shown in Fig. 29(e), and are measured to be $(1.21 \pm 0.03) \times 10^6$ and $(1.02 \pm 0.03) \times 10^6$ m/s, respectively. Additionally, the measured $v_e^F$ and $v_h^F$ are almost independent of the positions in the ripple. This large electron–hole asymmetry is attributed to the enhanced next-nearest neighbor hopping caused by the lattice deformation and curvature in the rippled region. By introducing a nonzero $t'$ in graphene, the Fermi velocity of the filled state (electrons) increases, and that of the empty state (holes) decreases, as shown in Figs. 29(c) and 29(e). A finite value of $t' = 0.16t$ well describes the observed electron–hole asymmetry in the ripple shown in Fig. 29(b). Here, the observed electron–hole asymmetry differs for different graphene ripples, and the estimated $t'$ ranges from ~$0.02t$ to ~$0.2t$ in the rippled regions in our experiment. Theoretically, the calculated $t'$ varies widely depending on the tight-binding parameterization [4]. Recently, the $t'$ in high-quality graphene was measured to be ~0.4 eV in a polarization-resolved magneto-Raman-scattering experiment [209] and was determined to be ~0.3 eV by quantum capacitance measurements [210]. In our experiment, $t'$ is simply a fitting parameter in the tight-binding model to account for the electron–hole asymmetry observed in the graphene ripples. The observed values, when compared with those estimated in theory and those measured in previous experiments, are reasonable.

We will demonstrate that in strained graphene, $v_e^F$ and $v_h^F$ in the graphene monolayer can be quite different around the charged defect. Figure 30(a) shows a representative STM image of a graphene sheet with high-density atomic-scale defects. A $\sqrt{3} \times \sqrt{3}\,R30°$ interference pattern, as shown in Fig. 30(b), can be observed around the defect [200]. The STS spectra, as shown in Fig. 30(c), show the existence of a resonance peak above the Dirac point associated with the defect. Figure 30(d) shows typical spectra recorded under different magnetic fields around the defects. Obviously, the observed Landau quantization exhibits large asymmetry between the empty state above the Dirac point and the filled state below the Dirac point, as shown in Fig. 30(e). According to Eq. (1), we determine $v_e^F$ and $v_h^F$ to be $(1.21 \pm 0.03) \times 10^6$ and $(1.02 \pm 0.03) \times 10^6$ m/s, respectively. Around the defects (within 10–15 nm), the electron–hole asymmetry is almost independent of the recorded positions, as shown in Fig. 31(a). Comparing the structure of the graphene sheet in Fig. 30(a) with that of the pristine graphene monolayer, the main differences are the existence of nanoscale rippling and the atomic defects. These nanoscale ripples, which usually occur with $h^2/(la) < 0.0254$, cannot induce such a large electron–hole asymmetry (here, $h$ is the amplitude, $l$ is the width of the ripple, and $a = 0.142$ nm), and more importantly, the nanoscales ripple with larger $h^2/(la)$—as shown in Fig. 29(b)—exhibit opposite electron–hole asymmetry to that observed around the defects (Fig. 30). Therefore, we mainly attribute the large electron–hole asymmetry shown in Fig. 30 to the charged-defect scattering. The electron-like and hole-like Dirac fermions in the graphene monolayer are expected to respond differently to a Coulomb potential: they are scattered more strongly when they are attracted to the charged defect than when they are repelled from it, as shown in Fig. 31(b) [211–213]. That is, charged Dirac fermions around an attractive potential center spend more time there and are more significantly deflected than those around a repulsive potential center. The attractive force and enhanced scattering induced by the charged defect can affect the transport properties of the quasiparticles and reduce the Fermi velocity of the quasiparticles. According to the observed electron–hole asymmetry, these defects shown in Fig. 30(a) are determined to have a positive charge, which is consistent with the fact that the resonance peak of the charged defect is located above the charge neutrality.

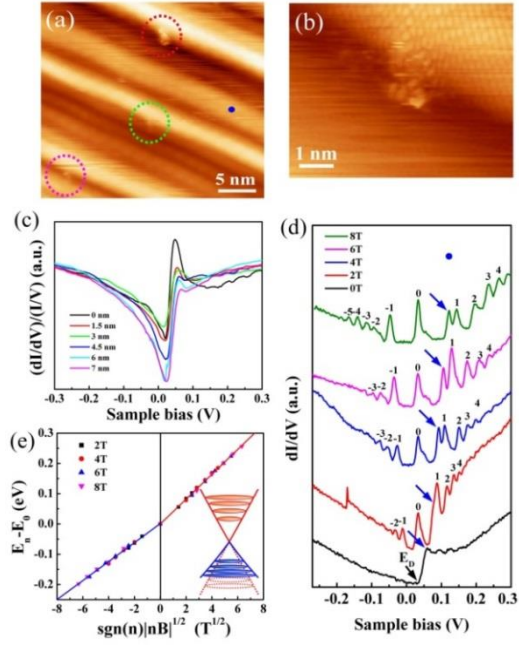

**Fig. 30** STM images and STS spectra of a graphene monolayer with a charged defect. **(a)** An STM topographic image of a graphene monolayer with three atomic defects (marked by dashed ellipses). **(b)** Magnified STM image of a defect in the red ellipse in (a). **(c)** Normalized STS spectra, *(dI/dV)/(I/V)-V*, measured on graphene at different distances from the center of the defect in (b). **(d)** *dI/dV* spectra taken under different fields *B* at the blue dot marked in (a). The blue arrows indicate the resonance peak of the defect. **(e)** LL peak energies of (d) plotted against sgn(n)|n|$B^{1/2}$. The red and blue solid lines show the linear fits of the data with Eq. (1) for holes and electrons, respectively. Inset: the schematic LL structure of the graphene sheet around (solid curves) and without (dotted curves) the charged defect. Reproduced from Ref. [76].

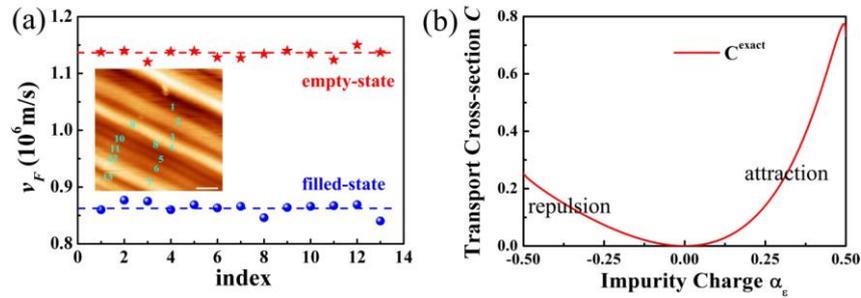

**Fig. 31 (a)** Summary of the Fermi velocities of electrons and holes obtained at different positions around the charged defects. **(b)** Transport cross-section **C** as a function of the impurity charge $\alpha_\varepsilon$. $\alpha_\varepsilon > 0$ indicates an attractive interaction between the impurity and the carriers, and $\alpha_\varepsilon < 0$ indicates a repulsive interaction between them. Reproduced from Ref. [76].

## 5.2 Valley-polarized LLs

In strained graphene, lattice deformation can create pseudo-magnetic fields affecting the behavior of massless Dirac fermions, resulting in zero-field LL-like quantization [199, 203–205, 214]. The primary difference between the strain-induced pseudo-magnetic field $B_S$ and the external magnetic field $B$ is that $B_S$ preserves time-reversal symmetry and has opposite signs in the two low-energy valleys of graphene: $K$ and $K'$ [215–221]. Therefore, the combination of the pseudo-magnetic field and the magnetic field can lift the valley degeneracy of the LLs and lead to unconventional valley-polarized Landau quantization in graphene monolayers [214]. In the experimental STM spectroscopic measurements, we directly observed valley-polarized LLs in strained graphene grown on Rh foil, which are induced by the coexistence of the pseudo-magnetic fields and external magnetic fields [214].

Strained structures can easily be observed for graphene grown on metallic substrates because of the mismatch of the thermal-expansion coefficients between graphene and the supporting substrates [198, 199, 204, 205, 222]. Figures 32(a) and 32(b) show representative STM images of a strained graphene structure on Rh foil. The STS spectra [Fig. 32(c)] recorded in the strained region under magnetic fields exhibit the Landau quantization of massless Dirac fermions in the graphene monolayer, with its characteristic non-equally spaced energy-level spectrum of LLs and the hallmark zero-energy state [45, 55, 57]. The linear fit of the experimental data to Eq. (1), as shown in Fig. 1(d), yields $v_F^e = (1.257 \pm 0.009) \times 10^6$ m/s and $v_F^h = (0.930 \pm 0.014) \times 10^6$ m/s. The large electron–hole asymmetry is mainly attributed to the enhanced next-nearest-neighbor hopping $t' \approx 0.3t$ caused by the lattice deformation and curvature in the strained graphene [4, 5, 204], as previously discussed.

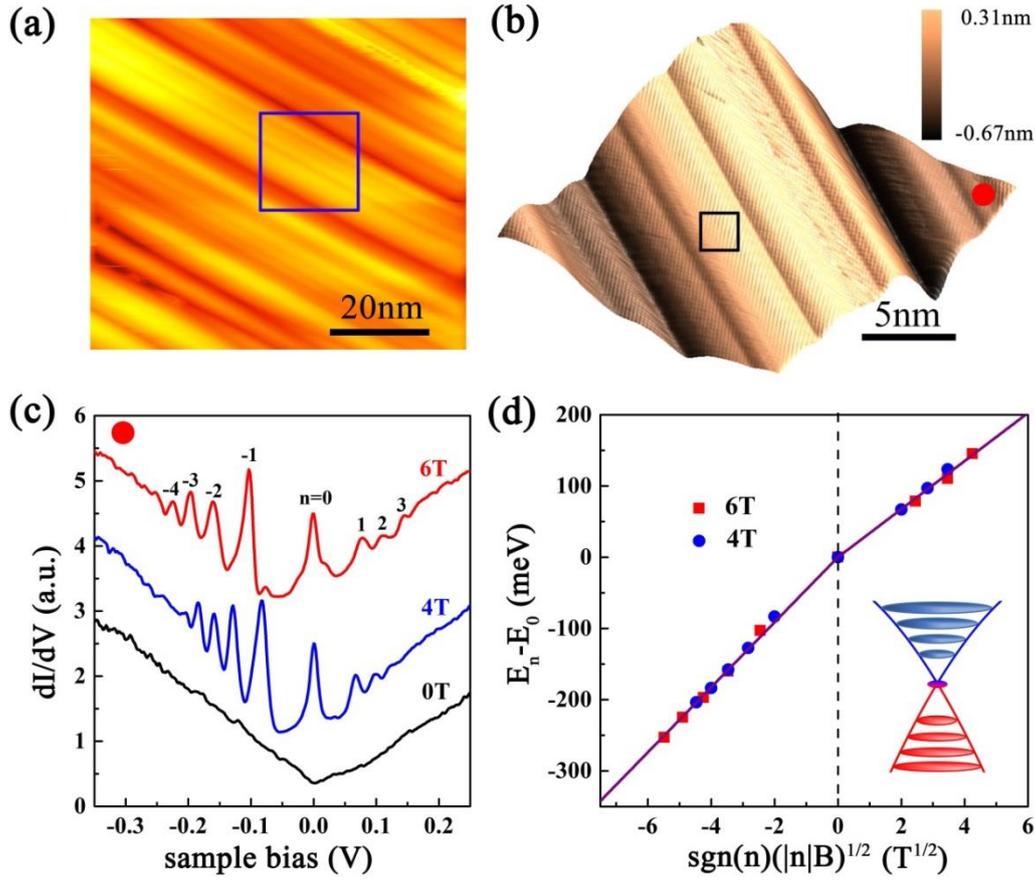

**Fig. 32** STM images and STS spectra of the strained graphene monolayer on Rh foil. **(a)** STM image of a strained graphene region showing 1D quasi-periodic ripples on Rh foil. **(b)** Enlarged STM image of the graphene ripples in the blue frame in (a). **(c)** STS spectra taken at the position marked with the red solid circle in (b) under different $B$. The spectra are shifted to maintain the $n = 0$ LL at the same bias. **(d)** LL peak energies obtained from (c) show a linear dependence on $\mathrm{sgn}(n)(|n|B)^{1/2}$, as expected for the graphene monolayer. Inset: schematic of the LLs in monolayer graphene, considering the next-nearest-neighbor hopping energy $t'$. Reproduced from Ref. [214].

In addition to the electron–hole asymmetry, we observe two other notable features in the STS spectra of some strained graphene regions, as shown in Fig. 33. One feature is the emergence of several peaks: a pronounced peak at the CNP and several weak peaks are observed at a relatively high bias, even in the spectra measured under zero magnetic field [Fig. 33(a)]. The other feature of the spectra is the splitting of $n = -1$ and $n = -2$ LLs recorded under different magnetic fields, as shown in Fig. 33(b). For the spectra recorded at the same position, the energy difference of the two peaks of $n = -1$ LL, $\Delta_{-1}$, decreases with increasing magnetic fields [Fig. 33(b)].

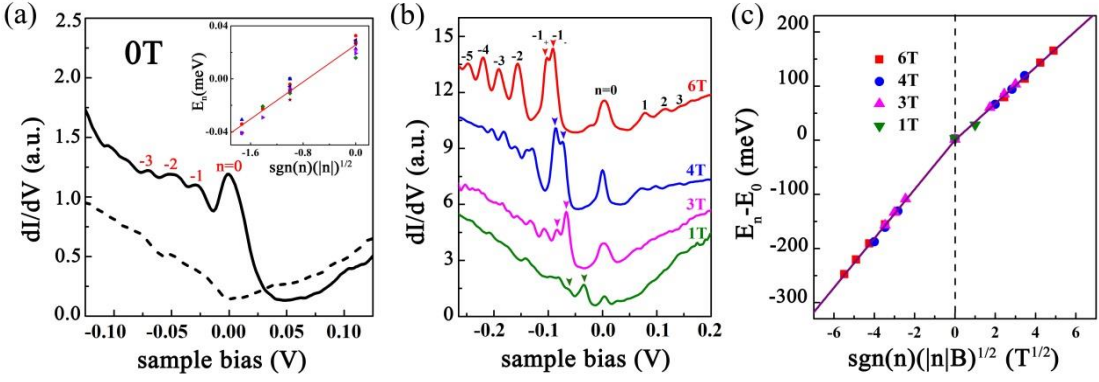

**Fig. 33** Pseudo-LLs and valley-polarized Landau quantization. **(a)** Two representative zero-field STS spectra recorded in graphene ripples. **(b)** STS spectra taken in the area marked with the black frame in Fig. 1(b) under different magnetic fields. The LL indices of massless Dirac fermions are marked. In the presence of an external $B$, the $n = -1$ LL splits into two peaks: $-1_-$ and $-1_+$ (here, the subscripts $-$ and $+$ denote $K$ and $K'$ valleys, respectively). **(c)** LL peak energies obtained from (a), showing a linear dependence on $\mathrm{sgn}(n)(|n|B)^{1/2}$. For clarity, the data for the split $n = -1$ LL are not plotted in this figure. The Fermi velocities for electrons and holes are measured to be $(1.250 \pm 0.007) \times 10^6$ and $(0.927 \pm 0.007) \times 10^6$ m/s, respectively. Reproduced from Ref. [214].

Our experimental observations can be understood within a theoretical framework that incorporates the effects of both strain-induced pseudo-magnetic fields and external magnetic fields on the Landau quantization of massless Dirac fermions in a graphene monolayer. In the case of $B = 0$ T, the pronounced peak at the CNP, as shown in Fig. 33(a), is attributed to the strain-induced partially flat bands ($n = 0$ LL) at zero energy [205, 223, 224], and the weak peaks at a high bias are attributed to higher pseudo-LLs. The signal of the higher pseudo-LLs is rather weak, which agrees with earlier STM measurements in strained graphene ripples [205]. This result is reasonable because the description of the hopping modulation in strained graphene as an effective pseudo-magnetic field is exactly valid only at the CNP, and the higher pseudo-LLs are less defined [223, 224]. From the energy spacing between the pseudo-LLs measured at $B = 0$ T, the pseudo-magnetic field is measured to be from 0.45 to 0.85 T in the studied ripples, and the average value is estimated to be $(0.60 \pm 0.05)$ T according to Eq. (1) (in calculating $B_S$, we use $v_F^e = (1.250 \pm 0.007) \times 10^6$ m/s, which was measured in our experiment).

Theoretically, a 2D strain field $u_{ij}(x,y)$ in graphene can induce a gauge field $A = \dfrac{\beta}{a}\begin{pmatrix} u_{xx} - u_{yy} \\ -2u_{xy} \end{pmatrix}$ and generate a pseudo-magnetic field $\vec{B}_S = \nabla \times A = -\dfrac{\beta}{a}\left(\dfrac{2\partial u_{xy}}{\partial x} + \dfrac{\partial(u_{xx} - u_{yy})}{\partial y}\right)$ (here, $a$ is on the order of the C-C bond length, and $2 < \beta = -\partial \ln t / \partial \ln a < 3$.) [145, 216, 225]. Obviously, not all the strained graphene structures result in a non-zero pseudo-magnetic field [226, 227], and it is difficult to generate a uniform pseudo-magnetic field in large-area graphene [216, 225]. Therefore, the observed pseudo-magnetic field varies spatially in our experiment, and only a part of the graphene ripples exhibit the pseudo-LLs. The pseudo-magnetic field arises from the spatial variation of the nearest-neighbor hopping $t$ in graphene [216, 225]. The results shown in Fig. 33 indicate that the local strain in the region where the spectra of Fig. 33 are recorded not only enhances the $t'$ but also results in the spatial variation of $t$. A local strain of 1% in graphene is predicted to generate a pseudo-magnetic field of 10 T [224]. Therefore, the observed $B_S \approx 0.6$ T is reasonable considering the width and height of the studied nanoscale ripples. In a previous study [204], the pseudo-magnetic field observed in graphene nanobubbles was as large as 300 T, indicating a local strain far larger than 10%. Such a large strain is attributed to the smaller width and larger height of the nanobubbles. The $h^2/(la)$ for the ripples studied in this work is far smaller than that of the nanobubbles.

In the presence of both the pseudo-magnetic field and the external magnetic field, the total effective magnetic field in one of the valleys, for example, the $K$ valley, is $B - B_S$, and that of the other valley, for example, the $K'$ valley, is $B + B_S$, as schematically shown in Fig. 34(a). Then, the valley degeneracy of the $n \neq 0$ LLs is expected to be lifted [8, 11] and the energy spacing of the two valleys for the $n_{th}$ LL is

$$\Delta_n = \sqrt{2e\hbar v_F^2 |n|(B+B_S)} - \sqrt{2e\hbar v_F^2 |n|(B-B_S)}, \quad n = ...-2, -1, 1, 2... \quad (10)$$

Considering that the picture of the effective pseudo-magnetic field is less defined at a high energy, the valley-polarized LLs should be observed only for small Landau indices $|n|$, for example, the $n = \pm 1$ LLs. In our experiment, the valley-polarized LL is clearly observed for $n = -1$, as shown in Fig. 33(b). Although the absence of splitting for the $n = 1$ LL measured at $B \neq 0$ T remains to be understood, it is likely that this observation is closely linked to the electron–hole asymmetry observed in our experiment. Additionally, the tunneling peak for the $n = 1$ LL is rather weak, which almost removes the possibility of detecting the valley splitting. Figure 34(b) summarizes the values of $\Delta_{-1}$ with respect to the external magnetic fields recorded at the same position of the strained graphene. The fit of the experimental result to Eq. (2) yields $B_S = (0.59 \pm 0.03)$ T, which is well consistent with that estimated according to the energy spacing of pseudo-LLs measured under zero magnetic field. The magnetic-field dependent of the splitting [Fig. 34(b)] provides convincing evidence for the valley-polarized LL induced by the external magnetic field and the pseudo-magnetic field.

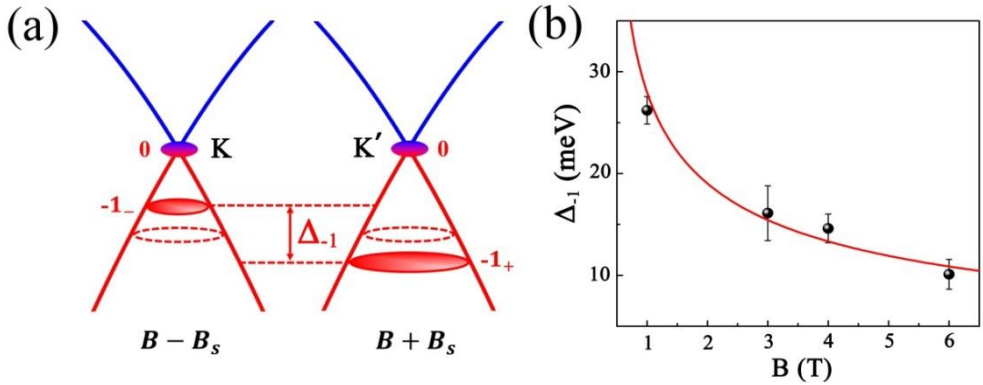

**Fig. 34** Valley-polarized LL and pseudo-magnetic fields in the strained graphene. **(a)** Schematic image showing Landau quantization in the graphene monolayer in the presence of both the pseudo-$B$ and the external $B$. The valley degeneracy of the $n = -1$ LL is lifted, and the energy spacing of the valley-polarized LLs is described by Eq. (10). **(b)** This figure summarizes the $\Delta_{-1}$ measured at the same position as a function of the external magnetic field. The solid curve is a fit to Eq. (10), with $B_S = (0.59 \pm 0.03)$ T as the only fitting parameter. Reproduced from Ref. [214].

## 6  Conclusions and perspectives

We reviewed recent experimental progress on Landau quantization in graphene and its multilayers using STM and STS. As we showed, by performing LL spectroscopic measurements of graphene sheets, the number of layers and the interlayer stacking configurations can be unambiguously verified. Compared with other conventional 2D electron systems, graphene materials offer a strictly 2D platform with fully exposed electronic states, which guarantees direct observations of long-desired quantum phenomena such as the two-component nature of Dirac quasiparticles and the LL bending via high-energy and spatial-resolution LL spectroscopy, as discussed in this review. Moreover, unusual Landau quantized behaviors in strained and defective graphene have been described, including the giant electron–hole asymmetry generated by strain and charged-defect scattering and the valley-polarized splitting of LLs in the presence of both a pseudo-magnetic field and a real field.

As documented in this review, the aforementioned fundamentals, as well as other insights, are important for understanding new quantum Hall phases in graphene and may help in designing future experiments to detect other physical phenomena. For example, i) it is expected that the magnetic fields can generate a variety of new interaction driven states, with natures that are rather different from those found in the conventional 2DEGs. Over the past decades, multicomponent FQHE have been experimentally observed in graphene. However, the sequence of FQHE fractions and their origin are poorly understood. To solve these problems, high-resolution tunneling spectroscopic measurements at a super-low temperature are an alternative method. ii) Owing to the extra degree of freedom provided by the layer, an abundance of exotic quantum Hall edge states is believed to emerge in different stacked graphene multilayers. Through using LL spectra with both an atomically spatial resolution and a high-energy resolution, the quantum Hall edge physics can be tested systematically at the well-arranged graphene edges under super-low potential fluctuations. iii) The recent studies on layer-stacking domain-wall solitons are very exciting and vibrant. The emergence of the topologically protected conducting channels under zero magnetic field makes such strain-like solitons promising candidates for ultralow power-consumption nanodevice applications. We consider that with in-situ identification of the stacking domain walls by STM, deeper insights and the complete physics of these confined electronic systems will be revealed. iv) As we demonstrated, the combination of the external real magnetic fields and the pseudo-magnetic fields induced by the spatially varying strain in a graphene monolayer leads to unconventional splitting of LLs not present in graphene under either field alone. We envision probing more exotic phenomena if the real magnetic fields and the pseudo-magnetic fields are combined with other properties of the graphene monolayer and the additional degrees of freedom found in graphene bilayers and multilayers.

**Acknowledgements** This work was supported by the National Natural Science Foundation of China (Grant Nos. 11674029, 11422430, 11374035), the National Basic Research Program of China (Grants Nos. 2014CB920903, 2013CBA01603), and the program for New Century Excellent Talents in University of the Ministry of Education of China (Grant No. NCET-13-0054). L.H. also acknowledges support from the National Program for Support of Top-notch Young Professionals.